\documentclass[aps,pre,reprint,onecolumn,showpacs]{revtex4}
\usepackage{amsmath}
\usepackage{epsfig}
\usepackage{amssymb,graphicx}
\usepackage{hyperref}
\usepackage{fancyhdr}
\newfont{\logo}{logo10}

\newcommand{\bea}{\begin{eqnarray}}
\newcommand{\eea}{\end{eqnarray}}
\newcommand{\bes}{\begin{subequations}}
\newcommand{\ees}{\end{subequations}}
\newcommand{\ds}{\displaystyle}

\renewcommand{\thispagestyle}[1]{}

\begin{document}

\title{Multicomponent long-wave--short-wave resonance interaction system:\\ Bright solitons, energy-sharing collisions, and resonant solitons}
\author{K. Sakkaravarthi}\email[ ]{ksakkaravarthi@gmail.com}
\affiliation{Post Graduate and Research Department of Physics, Bishop Heber College, Tiruchirappalli--620 017, Tamil Nadu, India}
\author{T. Kanna}\email[Corresponding author: ]{kanna\_phy@bhc.edu.in, kanna.phy@gmail.com}
\affiliation{Post Graduate and Research Department of Physics, Bishop Heber College, Tiruchirappalli--620 017, Tamil Nadu, India}
\author{M. Vijayajayanthi}\email[ ]{vijayajayanthi.cnld@gmail.com}
\affiliation{Department of Physics, Anna University, Chennai--600 025, Tamil Nadu, India}
\author{M. Lakshmanan}\email[ ]{lakshman@cnld.bdu.ac.in}
\affiliation{Centre for Nonlinear Dynamics, School of Physics, Bharathidasan University, Tiruchirapalli--620 024, Tamil Nadu, India}

%\date{First version: 8 May 2014, Final version: 10 November 2014}

\begin{abstract}
We consider a general multicomponent (2+1)-dimensional long-wave--short-wave resonance interaction (LSRI) system with arbitrary nonlinearity coefficients, which describes the nonlinear resonance interaction of multiple short waves with a long-wave in two spatial dimensions. The general multicomponent LSRI system is shown to be integrable by performing the Painlev\'e analysis. Then we construct the exact bright multi-soliton solutions by applying the Hirota's bilinearization method and study the propagation and collision dynamics of bright solitons in detail. Particularly, we investigate the head-on and overtaking collisions of bright solitons and explore two types of energy-sharing collisions as well as standard elastic collision. We have also corroborated the obtained analytical one-soliton solution by direct numerical simulation. Also, we discuss the formation and dynamics of resonant solitons. Interestingly, we demonstrate the formation of resonant solitons admitting breather-like (localized periodic pulse train) structure and also large amplitude localized structures akin to rogue waves coexisting with solitons. For completeness, we have also obtained dark one- and two-soliton solutions and studied their dynamics briefly.
\end{abstract}

% \newline  \newline Journal reference: {\it Phys. Rev. E} {\bf 90}, 052908 (2014)}
\maketitle

\section{Introduction}
Nonlinear wave interactions in physical systems lead to the formation of special nonlinear waves like solitons/soliton-like structures, shock waves, rogue waves, vortex solitons and so on \cite{Whitham-book,Akm-book}. The appearance of such nonlinear waves in almost every physical system motivated researchers from various disciplines of science to investigate their underlying remarkable dynamical behaviour in order to unearth non-trivial dynamical properties. Long-wave--short-wave resonance interaction (LSRI) is an interesting nonlinear interaction phenomenon that finds diversified applications namely in water waves, plasma physics, nonlinear optics, bio-physics and Bose-Einstein condensates \cite{lsri-opt,lsri-bec,Davydov}. This LSRI process arises during the nonlinear interaction between low- frequency long waves (LWs) and high-frequency short waves (SWs). In fact, there occurs a resonance interaction between the long wave and short waves when the phase velocity of the LW matches exactly/approximately the group velocity of the SWs. The pioneering study of such resonant nonlinear wave interaction in the context of plasma physics was made by Zakharov \cite{Zakh1972}. Kawahara {\it et al.} have investigated the nonlinear interaction between short- and long-capillary gravity waves \cite{Kawahara1975a}. The energy exchange between a nonlinear electron-plasma wave and a nonlinear ion-acoustic wave through resonance interaction mechanism was studied in Ref. \cite{Nishikawa1974}. At the same time, the LSRI phenomenon has been investigated independently by Benny \cite{Benny} and by Yajima and Oikawa \cite{Oikawa1976ptp} to study the interaction of ion sound wave with the Langmuir wave. Since then, several theoretical and experimental works have been reported based on the LSRI phenomenon in various contexts.

The LSRI phenomenon arising due to the interaction between long gravity wave and capillary-gravity wave for finite-depth water was investigated in Ref. \cite{Djor1977} by deriving a model equation and the solutions of the model equation were obtained \cite{Ma1978}. Later on, the resonant interaction of long and short internal waves in a three-layer fluid was studied experimentally \cite{Kopp1981}. Apart from this, study on the resonant coupling between ultra-long equatorial wave and packets of short gravity wave was also carried out in Ref. \cite{Boyd1983}. By using perturbation method, the one- and two-dimensional LSRI equations were obtained in Refs. \cite{Funakoshi1983,Funakoshi1989} for a two-layer fluid model and soliton (bright and dark type) solutions were constructed by applying the Hirota method \cite{Funakoshi1989}.

The two-dimensional analog of two-component LSRI system was investigated by Ohta {\it et al.} \cite{Ohta2007jpa}, in which they have derived the governing equation for a physical setting describing the interaction of nonlinear dispersive waves of three channels. Also, they have obtained special multi-soliton solutions using the Hirota method in the Wronskian form and analyzed their interactions \cite{Ohta2007jpa}. The Painlev\'e analysis of the two-component LSRI equation studied by Ohta {\it et al.}, has been carried out in \cite{Radha2009jpa} and there itself special dromion solutions have been obtained using the truncated Painlev\'e approach. Recently, the present authors have obtained more general multi-soliton solutions displaying a fascinating energy sharing (shape changing) collision in two-dimensional multicomponent LSRI equation \cite{Kanna2009jpa}. Very recently, the dynamics of bright soliton bound states of (2+1)-dimensional multicomponent LSRI system is investigated in detail in Ref. \cite{Sakkara2013epjst}.

Generalizing the procedure given in Ref. \cite{Ohta2007jpa} for three nonlinear dispersive waves, the propagation equation for multiple dispersive waves (say ($M+1$) waves) in a weak Kerr type nonlinear medium in the small amplitude limit can be obtained as shown in Ref. \cite{Kanna2012arxiv}. The corresponding set of general non-dimensional multicomponent (2+1)D LSRI system governing the resonance interaction between multiple SWs (say $M$) with a LW is given by
\bes\bea
&&i (S_{t}^{(\ell)}+\delta^{(\ell)} S_{y}^{(\ell)})- S_{xx}^{(\ell)}+ L S^{(\ell)}=0, \qquad \ell=1,2,3,...,M, \\
&&L_{t}=2\sum_{\ell=1}^M c_{\ell} |S^{(\ell)}|^2_{x},
\eea\label{model}\ees
where $S^{(\ell)}$ represents the $\ell$-th SW, $L$ indicates the LW and the subscripts represent the partial derivatives with respect to the evolutional coordinate $t$ and the spatial coordinates ($y$ and $x$). In the above (2+1)D LSRI system, ``2" stands for the two spatial dimensions ($x$ and $y$) and ``1" stands for the evolutional coordinate ``$t$". It should be noticed that in the waveguide geometry $t$ is the propagation direction, $x$ and $y$ denote the transverse coordinates. Here, $\delta^{(\ell)}$ and $c_{\ell}$ are real arbitrary parameters that determine the nature of higher dimensionality and the strength of nonlinear coupling of SWs, respectively. The nonlinearity coefficients $c_{\ell}$ in Eq. (\ref{model}b), in particular, their signs play pivotal role in determining the dynamics of system. Physically, these coefficients $c_{\ell}$ can be related to the self-phase modulation (SPM) and cross-phase modulation (XPM) coefficients in the context of nonlinear optics. Especially in the $j$-th SW component, the nonlinearity coefficients $c_{\ell}$ with $\ell = j$ ($\ell \neq j$) correspond to SPM (XPM) coefficients. In this work, we have clearly brought out the role of this parameter on the propagation and collision dynamics of bright solitons.

Interestingly, the one-component version ($M=1$) of the above system (\ref{model}) can be derived from a set of two-dimensional coupled nonlinear Schr\"odinger equations \cite{Onorato-prl,Shukla-prl}, describing the interaction of two-dimensional two waves propagating in different wave directions, when long-wave--short-wave resonance takes place, by using the approach of \cite{Ohta2007jpa}. This clearly indicates that system (\ref{model}) is a natural generalization of the two-wave system to the ($M+1$)-wave system in (2+1)-dimensions.

Very recently, we have derived the general ($1+1$)-dimensional multicomponent Yajima-Oikawa system from a set of multiple coupled nonlinear Schr\"odinger equations when LSRI takes place and have shown that the system is integrable via Painlev\'e test \cite{Kanna2013pre}. In addition to this, the bright multisoliton solutions and their interesting dynamics have been analyzed in Ref. \cite{Kanna2013pre}. The present system is a two dimensional generalization of the multicomponent Yajima-Oikawa system with a change of sign before the second derivative term in (\ref{model}a) and is of considerable physical importance. This clearly shows the physical significance of the considered system (\ref{model}).

Additionally, motivated by the intriguing collision scenario of bright solitons in integrable multicomponent nonlinear systems with mixed type (focusing-defocusing) nonlinearities \cite{Kannamixed} we wish to explore such special features in the present system too. For this purpose, first we study the integrability property of the above general $M$-component LSRI system. To the best of our knowledge, only a sub-system of (\ref{model}) that can be reduced for $M=2$, $c_1=c_2=1$ and $\delta^{(1)}=\delta^{(2)}=1$ has been shown to be integrable by Painlev\'e analysis \cite{Radha2009jpa}. However, the integrability aspects of the present general (2+1)D $M$-LSRI system (\ref{model}) has not been investigated so far. Then we construct the multi-soliton solutions in Gram determinant form using Hirota's method and analyze their different types of energy sharing interactions.

An interesting aspect of soliton studies is the formation and propagation of resonant solitons, which arise as a special case of multi-soliton solutions. The resonant soliton (RS) occurs when the phase-shift experienced by the colliding solitons becomes infinity or tends to become infinity \cite{reso-ref,reso-main,reso-ref2}. The difference between the resonance mechanism in LSRI process and in the RS is that the former arises when the group and phase velocities of the interacting nonlinear waves match each other while the latter appears for particular choice of soliton solution parameters, that is, the choice of the soliton parameters for which the phase-shift of the colliding waves (solitons) becomes infinity. For the RS, the amplitude reaches a maximum value and the nature of soliton energy switching becomes non-trivial. In general, RSs appear in integrable higher dimensional nonlinear systems which admit multi-soliton solution, as these RSs arise only as special cases of multi-soliton solutions describing the interaction of multiple solitons. These resonant solitons produce different types of interaction patterns such as Y-type, inverted Y-type and coupled (Y--inverted Y) type structures which may find applications in junction couplers. Especially, in the case of two-soliton resonance, the two colliding solitons may combine into a single soliton after collision (soliton fusion) or a single soliton may split up into two solitons after collision (soliton fission) or the two solitons collide, travel like a single soliton for certain distance and then separate into individual solitons (long-range interaction) \cite{reso-1d,reso-main}. So, it is our further interest to investigate the formation of such resonant solitons and their subsequent dynamics in the present system (\ref{model}). Also, we construct the dark one- and two- soliton solutions of $M$-LSRI system (\ref{model}) and briefly discuss their propagation and collision dynamics for completeness. These dark solitons are interesting nonlinear objects which occur for asymptotically non-vanishing boundary conditions \cite{Kiv1993ol,Kiv1997pre,Ohtadark}.

The rest of the article is organized as follows. In Sec. \ref{secsoli}, we briefly outline the integrability nature of the system by applying the Painlev\'e analysis and obtain the bright multi-soliton solutions by using the Hirota's direct method \cite{Hirota-book}. The dynamics of bright one-soliton is explained in Sec. \ref{seconesol} and two different types of energy sharing collisions of two bright solitons are discussed in Sec. \ref{collisions}. In Sec. \ref{secreso}, we unearth the features of resonant solitons of $M$-LSRI system (\ref{model}). The dark soliton solutions of system (\ref{model}) are given in Sec. \ref{secdark} and the results are summarized in the final section.

\section{Painlev\'e analysis and Multi-soliton solutions}\label{secsoli}
The Painlev\'e analysis of Eq. (\ref{model}) can be carried out in a standard way \cite{Weiss,painml} as done in Ref. \cite{Kanna2013pre} for the one-dimensional $M$-component Yajima-Oikawa system, with arbitrary $M$. For completeness, we briefly outline the main steps of the Painlev\'e analysis of system (\ref{model}) in the Appendix. There we show that in fact the $M$-LSRI system (\ref{model}) admits ($2M+1$) number of integer resonances and also admits sufficient number of arbitrary functions at each of those ($2M+1$) resonances for arbitrary values of the real quantities $c_{\ell}$ with  $\delta^{(1)}=\delta^{(2)}=...=\delta^{(M)}$ ($\equiv \delta$, $\delta$ is a real constant, either positive or negative). Thus one can conclude that the $M$-LSRI system (1) is integrable in Painlev\'e sense for arbitrary real nonlinearity coefficients $c_{\ell}$ with $\delta^{(\ell)}$'s being the same. The role of these nonlinearity coefficients in the formation and dynamics of solitons can be revealed by constructing the explicit multi-soliton solutions of (\ref{model}) and by analyzing their underlying dynamics.

For this purpose, we obtain the bright multi-soliton solutions of the general $M$-LSRI system (\ref{model}), with $\delta^{(\ell)}=\delta$, by applying Hirota's  bilinearization method \cite{Hirota-book}. Equation (\ref{model}) can be expressed in the following bilinear form:
\bes\bea
&&\left(i(D_t+\delta D_y)-D_x^2\right)(g^{(\ell)} \cdot f)=0,\quad \quad \ell=1,2,3,...,M,\\
&&(D_xD_t-2\lambda) (f \cdot f)=-2\sum_{\ell=1}^{M} c_{\ell} |g^{(\ell)}|^2,
\eea\label{beq}\ees
by introducing the bilinearizing transformations to the dependent variables as $S^{(\ell)}=\frac{g^{(\ell)}}{f},\quad \ell=1,2,...M,$ and $L=-2\frac{\partial^2}{\partial x^2}(\ln{f})$ in Eq. (\ref{model}). Here $g^{(\ell)}$ and $f$ are arbitrary complex and real functions, respectively, $\lambda$ is a unknown constant and the Hirota's $D$-operators $D_j$, $j=x,t,y$, are defined as $D_x^{p}D_t^{q}D_y^{r}(a\cdot b) =\big(\frac{\partial}{\partial x}-\frac{\partial}{\partial x'}\big)^p\big(\frac{\partial}{\partial t}-\frac{\partial}{\partial t'}\big)^q \big(\frac{\partial}{\partial y}-\frac{\partial}{\partial y'}\big)^r a(x,t,y)b(x',t',y')\big|_{ \ds (x=x', t=t', y=y')}$\cite{Hirota-book}. For $\lambda=0$, Eqs. (\ref{beq}) admit bright soliton solutions with zero background, while for the general case, nonzero $\lambda$, (\ref{beq}) can admit bright-dark and dark-dark soliton solutions.
In the present work, we restrict our study to a detailed investigation on the bright soliton dynamics and a brief discussion on the dark-dark soliton solutions. The results on the mixed (bright-dark) soliton solutions of $M$-LSRI system (1) will be published as a separate paper.
%However, the present work is restricted to the  investigation of bright soliton dynamics and a detailed study on bright-dark and dark-dark solitons in system (\ref{model}) will be published as a separate paper.

The explicit form of bright $n$-soliton solution, for arbitrary $n$, is obtained by expressing the dependent variables $g^{(\ell)}$ and $f$ in terms of power series expansions as $g^{(\ell)}=\ds\sum_{j=1}^{n}\chi^{2j-1} g_{2j-1}^{(\ell)}$ and $f=1+\ds\sum_{j=1}^{n}\chi^{2j} f_{2j}$, respectively, and by recursively solving the equations resulting from Eq. (\ref{beq}) at different powers of $\chi$. We express the obtained bright $n$-soliton solution in Gram determinant form as below:
\bes\bea
S^{(\ell)}&=&\frac{g^{(\ell)}}{f}, \quad \ell=1, 2, \ldots, M,\\
L&=&-2\frac{\partial^2}{\partial x^2}(\ln{f}),%(\ln{f})_{xx},
\eea
where
\bea
g^{(\ell)}=
\left|
\begin{array}{ccc}
A & I & \phi\\
-I & B & {\bf 0}^T\\
{\bf 0} & a_{\ell} & 0
\end{array}
\right|, \quad \quad f= \left|
\begin{array}{cc}
A & I\\
-I & B
\end{array}
\right|.
\eea
In Eq. (\ref{nsol}c), $I$ and $\bf{0}$ represent identity matrix and null matrix of dimensions ($n \times n$) and ($1 \times n$), respectively, $A$ and $B$ are square matrices of dimension ($n \times n$) with elements
\bea
A_{ij}&=& \frac{e^{\eta_{i}+\eta_{j}^*}}{k_{i}+k_{j}^*},\\
B_{ij}&=&\kappa_{ji}=\frac{-\psi_i^{\dagger} c\psi_j}{(\omega_i^*+\omega_j)} \equiv \frac{-\sum_{\ell=1}^M c_{\ell}\alpha_j^{(\ell)} \alpha_i^{(\ell)*}}{(\omega_i^*+\omega_j)},\quad i,j=1, 2, \ldots, n.
\eea\label{nsol}\ees
The block-matrices $a_{\ell}$, $\psi_j$, $\phi$ and $c$ are of dimensions ($1\times M$), ($M \times 1$), ($n \times 1$) and ($M \times M$), respectively and are defined as $a_{\ell} = -\left(\alpha_1^{(\ell)}, \alpha_2^{(\ell)}, \ldots, \alpha_{n}^{(\ell)}\right)$, $\psi_j=\left(\alpha_j^{(1)},~\alpha_j^{(2)},\ldots,\alpha_j^{(M)}\right)^T$, $\phi = \left(e^{\eta_1},e^{\eta_2},\ldots,e^{\eta_{n}}\right)^T$ and $c= \mbox{diag}\left(c_1, c_2, \ldots, c_M \right)$, where $j=1, 2, \ldots,n$, $\ell=1,2,3,...,M$. Here $\eta_j=k_j x-\frac{1}{\delta}(ik_j^2+\omega_j)y+\omega_jt$, in which  $k_j$, $\omega_j$ and $\alpha_{j}^{(\ell)}$,  $j=1,2,\ldots,n$, $\ell=1,2,\ldots,M$, are arbitrary complex parameters. Throughout this paper, $M$ and $n$ represent the component number and soliton number, respectively, while the symbols $\dagger$ and $T$ appearing in the superscript indicate the transpose conjugate and transpose of the matrix, respectively. The proof of the above $n$-soliton solution (\ref{nsol}), which we have skipped here, can be easily done by expressing the bilinear equations (\ref{beq}), after substituting Eq. (\ref{nsol}), in the form of Jacobi identity (for details see Ref. \cite{Kanna2009jpa}).

The main difference between the soliton solution of present system to that of the solution given in Ref. \cite{Kanna2009jpa} is the quantity $B_{ij}$ and this plays vital role in defining the nature of soliton solution and in their collision dynamics as will be shown in the forthcoming sections. Particularly, the nature of solution (whether it is singular or non-singular) depends on the system parameters $c_{\ell}$, which appear in $B_{ij}$, in addition to the soliton parameters $\alpha_j^{(\ell)}$, $k_{j}$ and $\omega_{j}$. For various choices of $c_{\ell}$ parameters one can obtain solitons displaying different interesting dynamics. Hence the arbitrariness of nonlinearity coefficients $c_{\ell}$, particularly their signs, gives an additional freedom resulting in rich soliton dynamics.

\section{Bright one-soliton}\label{seconesol}
In this section, the dynamics of one soliton appearing for different choices of strength of nonlinearities is explored in detail. For this purpose, we write the explicit form of bright one-soliton solution of $M$-LSRI system (\ref{model}), resulting for the choice $n=1$ in Eq. (\ref{nsol}), as below:
\bes\bea
&&S^{(\ell)}=  A_{\ell} \sqrt{k_{1R}\omega_{1R}}~\mbox{sech} \left(\eta_{1R}+\frac{R}{2}\right) e^{i(\eta_{1I}-\frac{\pi}{2})}, \quad \ell=1,2,...M,\\
&&L=-2k_{1R}^2 \mbox{sech}^2 \left(\eta_{1R}+\frac{R}{2}\right),
\eea\label{1sol}\ees
where $A_{\ell}={\alpha_1^{(\ell)}}{\left(\sum_{\ell=1}^{M}c_{\ell} |\alpha_1^{(\ell)}|^2\right)^{-\frac{1}{2}}}$, $e^{R}=\frac{-\sum_{\ell=1}^{M} c_{\ell} |\alpha_1^{(\ell)}|^2}{4k_{1R}\omega_{1R}}$, $\eta_{1R}=k_{1R}x+\frac{1}{\delta}(2k_{1R}k_{1I}-\omega_{1R})y+\omega_{1R}t$ and $\eta_{1I}=k_{1I}x-\frac{1}{\delta}(k_{1R}^2-k_{1I}^2+\omega_{1I})y+\omega_{1I}t$. Here and in the following the subscripts $R$ and $I$ appearing in a particular complex parameter denote the real and imaginary parts of that complex parameter. The above bright one-soliton is characterized by ($M+2$) arbitrary complex parameters, $\alpha_1^{(\ell)}$, $k_1$ and $\omega_1$. In addition to these soliton parameters, one can also tune the system parameters, namely, the nonlinearity coefficients ($c_{\ell}$). The amplitudes (peak values) of soliton in the $\ell$-th SW component and LW component are $A_{\ell} \sqrt{k_{1R}\omega_{1R}}$ and $2k_{1R}^2$, respectively. One can observe that the amplitude of LW soliton is independent of $\alpha_1^{(\ell)}$-parameters and $\omega_{1}$. This shows the interesting possibility of controlling the SW soliton without altering the LW soliton by tuning $\alpha_1^{(\ell)}$ and $\omega_{1}$. It should be noticed that, in the present system the so-called line-solitons can propagate in two different planes, namely ($x-y$) plane and ($x-t$) plane for fixed $t$ and $y$, respectively. The velocity of soliton in the ($x-y$) plane is $\frac{1}{\delta}(\frac{\omega_{1R}}{k_{1R}}-2k_{1I})$ while the soliton velocity in ($x-t$) plane is $-\frac{\omega_{1R}}{k_{1R}}$. One can also notice that the velocity of propagating soliton in the ($x-y$) plane can be altered without affecting the velocity of soliton in the ($x-t$) plane by mere tuning of the $k_{1I}$ parameter. We would like to remark that the higher dimensionality coefficient $\delta$ especially affects the velocity of solitons in the ($x-y$) plane and shifts the position of solitons in the ($x-t$) plane. Here, in this paper, we investigate and explore several interesting points resulting from the arbitrariness of nonlinearity coefficients ($c_{\ell}$) for fixed $\delta$ parameter (say $\delta=1$).

One can observe that the nature of above one-soliton solution is determined by the quantity $e^R$. Particularly, singular solutions result for $e^R<0$ and non-singular solutions result for $e^R>0$. Such dependence of the existence of regular soliton on $e^R$ which in turn depends on the arbitrariness of $c_{\ell}$, leads us to classify the soliton solutions of $M$-LSRI system (\ref{model}) into three cases, (i) positive nonlinearity coefficients ($c_{\ell}>0$), (ii) negative nonlinearity coefficients ($c_{\ell}<0)$ and (iii) mixed-type coefficients (both positive and negative values of $c_{\ell}$), as in table \ref{table1}.

\begin{figure}[h]
\centering\includegraphics[width=0.9\linewidth]{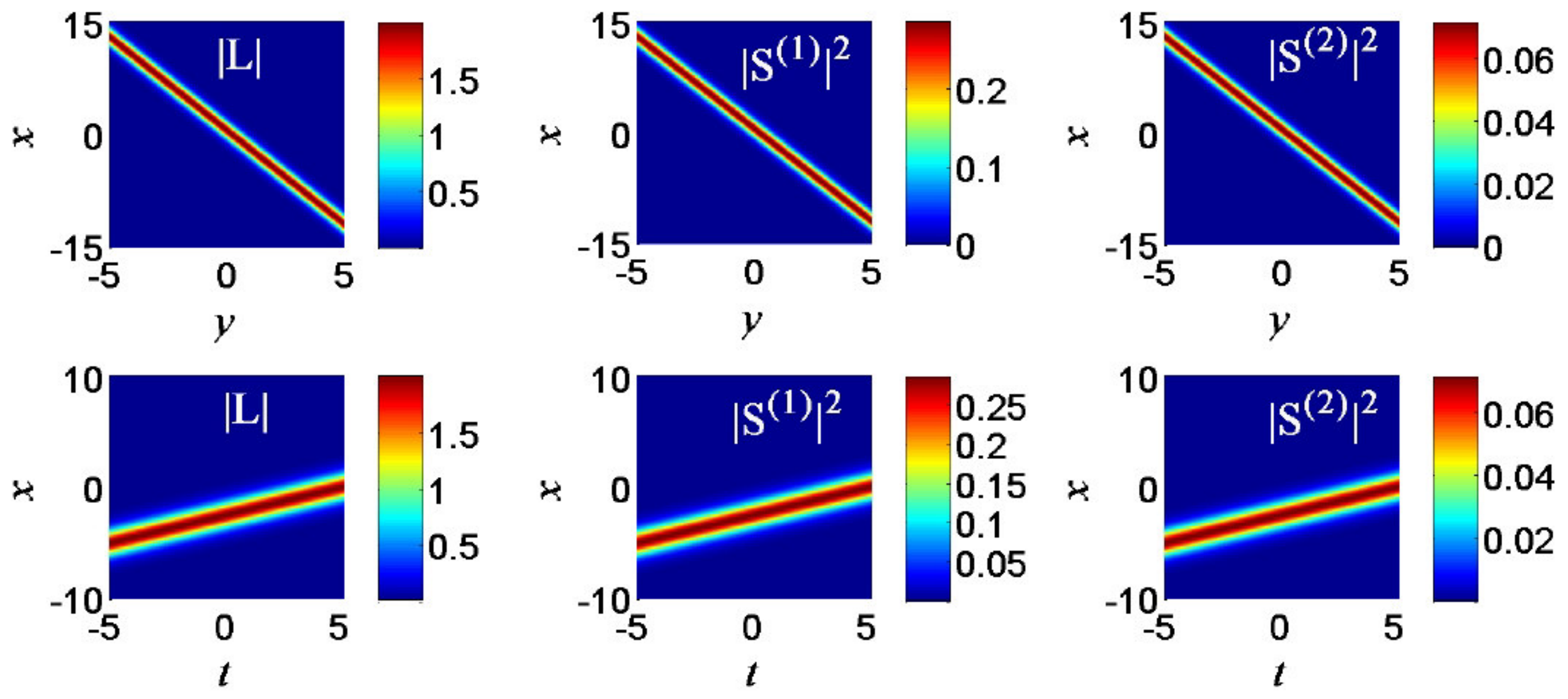}%a.eps}\includegraphics[width=0.31\linewidth]{figure1b.eps}\includegraphics[width=0.31\linewidth]{figure1c.eps}\\
%\centering\includegraphics[width=0.31\linewidth]{figure1d.eps}\includegraphics[width=0.31\linewidth]{figure1e.eps}\includegraphics[width=0.31\linewidth]{figure1f.eps}
\caption{(Color online) Propagation of bright one-soliton of 2-LSRI system with positive nonlinearity coefficients ($c_{\ell}>0$) in ($x-y$) plane for $t=1$ (top panels) and in ($x-t$) plane for $y=1$ (bottom panels). The parametric choice is $k_1=1+i$, $\omega_1=-0.5+i$, $c_1=1.5$, $c_2=1$, $\alpha_1^{(1)}=1$ and $\alpha_1^{(2)}=0.5$.}
\label{osfig1}
\end{figure}

\begin{table}[h]
{\scriptsize
%\begin{center}
\caption{Different choices of parameters for the existence of regular bright one-soliton in $M$-LSRI system (\ref{model})}
\label{table1}
\begin{tabular}{|c|p{4cm}|p{1.7cm}|p{3.2cm}|c|c|c|p{1.3cm}|}
  \hline\hline
  Case & ~~~~~~~~~Choice of $c_{\ell}$ & Choice of $\Lambda$ & ~~~Condition for &  \multicolumn{2}{|c|}{Amplitude of soliton in} & \multicolumn{2}{|c|}{Velocity of LW/SW soliton} \\\cline{5-8}
   & & & ~~regular soliton & $\ell$-th SW comp. & LW comp. & in ($x-y$) & in ($x-t$) \\ \hline\hline
  (i) & $c_{\ell}>0$, $\ell=1,2,...,M$ & ~~~~$\Lambda <0$ & $k_{1R}>0$; $\omega_{1R}<0$ (or) \newline $k_{1R}<0$; $\omega_{1R}>0$ & $2\alpha_1^{(\ell)} \sqrt{\left|\frac{k_{1R} \omega_{1R}}{\Lambda}\right|}$ & $2k_{1R}^2$ & $-\left|\frac{\omega_{1R}}{k_{1R}}\right|-2k_{1I}$ & ~~~$\left|\frac{\omega_{1R}}{k_{1R}}\right|$  \\ \hline
 (ii) & $c_{\ell}<0$, $\ell=1,2,...,M$ & ~~~~$\Lambda >0$ & $k_{1R}>0$; $\omega_{1R}>0$ & $2 \alpha_1^{(\ell)}\sqrt{\left|\frac{k_{1R} \omega_{1R}}{\Lambda}\right|}$ & $2k_{1R}^2$ & $\left|\frac{\omega_{1R}}{k_{1R}}\right|-2k_{1I}$ & $-\left|\frac{\omega_{1R}}{k_{1R}}\right|$  \\ \hline
 (iii) & {$c_{\ell}>0$, $\ell=1,2,...,Q$, \newline $c_{\ell}<0$, $\ell=Q+1,Q+2,...,M$} & ~~~~$\Lambda <0$ \newline($|\Lambda_1|>|\Lambda_2|$)& (a) $k_{1R}>0$; $\omega_{1R}<0$ (or) \newline (b) $k_{1R}<0$; $\omega_{1R}>0$ & $2\alpha_1^{(\ell)} \sqrt{\left|\frac{k_{1R} \omega_{1R}}{\Lambda_1-\Lambda_2}\right|}$ & $2k_{1R}^2$  & $-\left|\frac{\omega_{1R}}{k_{1R}}\right|-2k_{1I}$ & ~~~$\left|\frac{\omega_{1R}}{k_{1R}}\right|$  \\\cline{3-8}
  &  & ~~~~$\Lambda >0$ \newline ($|\Lambda_1|<|\Lambda_2|$) & (c) $k_{1R}>0$; $\omega_{1R}>0$ & $2\alpha_1^{(\ell)} \sqrt{\left|\frac{k_{1R} \omega_{1R}}{\Lambda_1-\Lambda_2}\right|}$ & $2k_{1R}^2$  & $\left|\frac{\omega_{1R}}{k_{1R}}\right|-2k_{1I}$ & $-\left|\frac{\omega_{1R}}{k_{1R}}\right|$  \\ \hline\hline
\end{tabular}}
{\newline In the above table, $\Lambda= - \ds\sum_{\ell=1}^M c_{\ell}|\alpha_1^{(\ell)}|^2$, $\Lambda_1=-\ds\sum_{\ell=1}^Q |c_{\ell}|~|\alpha_1^{(\ell)}|^2$ and $\Lambda_2=\ds\sum_{\ell=Q+1}^M |c_{\ell}|~|\alpha_1^{(\ell)}|^2$.}
%\end{center}
\end{table}
When all the nonlinearity coefficients are positive ($c_{\ell}>0$), the numerator of the expression for $e^R$ (see below Eq. (\ref{1sol})) always takes negative values. Hence the regular (non-singular) soliton solution can be obtained for the choice either $k_{1R}>0$ with $\omega_{1R}<0$ or $k_{1R}<0$ with $\omega_{1R}>0$, only for which the condition for  non-singular solution (that is, $e^R>0$) is satisfied. In Fig. \ref{osfig1}, we have shown the bright soliton of 2-LSRI system (Eq. (\ref{model}) with $M=2$) propagating in the ($x-y$) plane for $t=1$ (top panels) and in the ($x-t$) plane for $y=1$ (bottom planes) for the choice $c_1,c_2>0$, $k_{1R}>0$ and $\omega_{1R} < 0$. It is evident from Fig. \ref{osfig1}, that the solitons propagate with different velocities in the ($x-y$) and ($x-t$) planes (respectively, $-2.5$ and $0.5$). The quantities appearing in all the figures of this paper are adimensional. Additionally, one can also obtain similar type of solitons for $k_{1R}<0$ and $\omega_{1R} > 0$.\\

We have performed a direct numerical simulation of system (\ref{model}) by using the split-step Crank-Nicolson method \cite{PM2009cpc,PM2012cpc} and plot the one-soliton propagation in Fig. 2 corresponding to the initial conditions of Fig. 1. Here we have considered the domain $x\in [-40,40]$ with $\Delta x=0.1$, $\Delta y=0.1$ and $\Delta t=0.01$ \footnote{The numerical simulation was carried out in collaboration with P. Muruganandam}. The numerical results well corroborate the analytical results. It is a straightforward task to extend the numerical analysis to multi-soliton solutions of $M$-LSRI system (\ref{model}).
\begin{figure}[h]
\centering\includegraphics[width=0.7\linewidth]{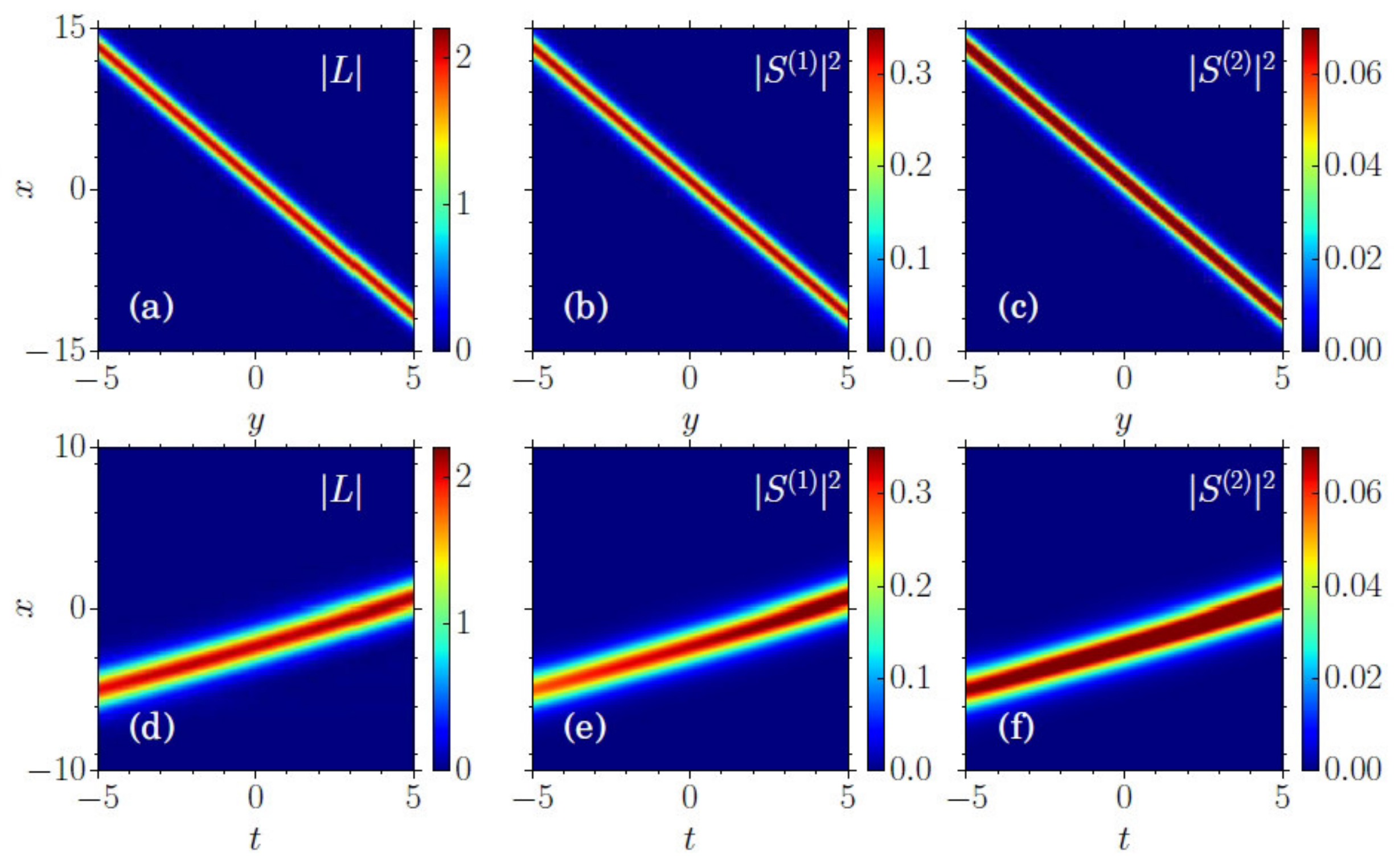}%1g.eps}
\caption{(Color online) Direct numerical simulation of bright one-soliton propagation in 2-LSRI system in ($x-y$) plane for $t=1$ (top panels) and in ($x-t$) plane for $y=1$ (bottom panels).}
\label{num-1s}
\end{figure}

For negative values of nonlinearity coefficients ($c_{\ell}<0$), the requirement for non-singular solution ($e^R>0)$ restricts $k_{1R}$ and $\omega_{1R}$ to be positive. It can be noticed that for a given set of parameters the amplitudes of the solitons in this case are the same as those of the previous case ($c_{\ell}>0$). But there occurs a significant change in the velocity of solitons in the ($x-y$) plane and the soliton velocity in the ($x-t$) plane now becomes opposite to that of the case $c_{\ell}>0$ (see Table \ref{table1}).

For mixed signs of nonlinearity coefficients (say, $c_{\ell}>0$ for $\ell=1,2,...,Q,$ and $c_{\ell}<0$ for $\ell=Q+1,Q+2,...,M$), all the soliton parameters ($k_{1R}$, $\omega_{1R}$ and $\alpha_j^{(\ell)}$) play crucial role in obtaining the non-singular soliton solution. We can have three different conditions (sub-cases) for achieving regular solutions: (a) $\Lambda<0$, $k_{1R}>0$, $\omega_{1R}<0$ (b) $\Lambda<0$, $k_{1R}<0$, $\omega_{1R}>0$, (c) $\Lambda>0$, $k_{1R}>0$, $\omega_{1R}>0$. Here the bright soliton in a particular SW component can have the same amplitude for the three subcases (a), (b) and (c), which is different from the previous two cases (i) and (ii), for a given set of soliton parameters ($k_1$, $\omega_1$ and $\alpha_1^{(\ell)}$) with fixed magnitude of $c_{\ell}$. However, the soliton velocities for the choices (a) and (b) in case (iii) are the same as that of case (i) while that of the choice (c) in case (iii) is the same as the velocity of case (ii) (see Table \ref{table1}).

The above analysis shows that the $c_{\ell}$ parameters can be profitably used for controlling the dynamics of solitons in the SW components in $M$-LSRI system (\ref{model}). It clearly indicates that the propagation characteristics of solitons can be tuned by suitably altering the system parameters $c_{\ell}$ which will find application in controlling soliton and also in pulse shaping in the context of nonlinear optics.

\section{Bright two soliton collisions}\label{collisions}
Bright two-soliton solution of $M$-LSRI system (\ref{model}), with $\delta=1$, can be obtained from Eq. (\ref{nsol}) by putting $n=2$. Explicitly the determinant forms of $g^{(\ell)}$ and $f$ can be written as
\bea
g^{(\ell)}=
\left|
\begin{array}{ccccc}
A_{11} & A_{12} & 1 & 0 & e^{\eta_1}\\
A_{21} & A_{22} & 0 & 1 & e^{\eta_2}\\
-1 & 0 & \kappa_{11} & \kappa_{21} & 0\\
0 & -1 & \kappa_{12} & \kappa_{22} & 0\\
0 & 0 & -\alpha_1^{(\ell)} & -\alpha_2^{(\ell)} & 0
\end{array}
\right|, \quad \quad f= \left|
\begin{array}{cccc}
A_{11} & A_{12} & 1 & 0 \\
A_{21} & A_{22} & 0 & 1 \\
-1 & 0 & \kappa_{11} & \kappa_{21}\\
0 & -1 & \kappa_{12} & \kappa_{22}\\
\end{array}
\right|,
\label{2sol}\eea
with $A_{ij}= \frac{e^{\eta_{i}+\eta_{j}^*}}{k_{i}+k_{j}^*}$, $\kappa_{ji}= \frac{-1}{(\omega_i^*+\omega_j)}\ds\sum_{\ell=1}^M c_{\ell}\alpha_j^{(\ell)} \alpha_i^{(\ell)*}$, $i,j=1, 2$, and $\eta_j=k_j x-(ik_j^2+\omega_j)y+\omega_jt$, $j=1,2$. The multicomponent nature of the system results in fascinating collision dynamics of solitons. The arbitrariness of the nonlinearity coefficients $c_{\ell}$, particularly, their signs play pivotal role in the collision dynamics displayed by the bright solitons in $M$-LSRI system, as mentioned in the introduction. The soliton collisions can be well understood by performing an asymptotic analysis \cite{RK1997pre,Kannamixed,Kanna-cnls,Kanna2009jpa,Kanna-ccnls} and we do not present the corresponding mathematical details here, for brevity. It is instructive to mention that the present system results in different collisional behavior in the ($x-y$) and also in the ($x-t$) planes as the solitons admit different velocities in those planes as will be shown below.

A careful asymptotic analysis shows that the change in the amplitude of a given soliton (say $j$-th soliton) after collision can be related to that of before collision, in the $\ell$-th SW component through the relation
\bea
A_j^{(\ell)+} = T_j^{(\ell)} A_j^{(\ell)-}, \qquad j=1,2, \qquad \ell=1,2,...,M,\label{asymp}
\eea
where the transition amplitudes $T_j^{(\ell)}$'s are defined as $T_1^{(\ell)} = \frac{1-\lambda_1}{\sqrt{1-\lambda_1 \lambda_2}}\left(\frac{(k_1-k_2)(k_2+k_1^*)}{(k_1^*-k_2^*)(k_2^*+k_1)} \right)^{1/2}$ and $T_2^{(\ell)} = \frac{\sqrt{1-\lambda_1 \lambda_2}}{1-\lambda_2} \left(\frac{(k_2+k_1^*)(k_1^*-k_2^*)}{(k_2^*+k_1)(k_1-k_2)} \right)^{1/2}$.
Here $\lambda_1=\frac{\alpha_2^{(\ell)} \kappa_{12}}{\alpha_1^{(\ell)} \kappa_{22}}$ and $\lambda_2=\frac{\alpha_1^{(\ell)} \kappa_{21}}{\alpha_2^{(\ell)} \kappa_{11}}$, where $\kappa_{ij},~i,j=1,2$, takes the form as given below Eq. (\ref{2sol}). When the transition amplitudes become unimodular (that is, $|T_j^{(\ell)}|^2=1$) there occurs an elastic collision. This is possible only for the choice $\frac{\alpha_1^{(1)}}{\alpha_2^{(1)}} = \frac{\alpha_1^{(2)}}{\alpha_2^{(2)}}= ... =\frac{\alpha_1^{(M)}}{\alpha_2^{(M)}}$. However, in a general setting, the bright solitons undergo energy sharing (shape-changing or energy exchange) collisions as $|T_j^{(\ell)}|^2 \neq1$. Importantly, for the solitons in system (\ref{model}), the nature of energy switching is determined by the strength (sign) of the nonlinearity coefficients ($c_{\ell}$) in addition to the soliton parameters. Additionally, the solitons (say $s_1$ and $s_2$) experience a phase-shift ($\Phi_1$ and $\Phi_2$) given by
\bea
\Phi_1&=&\frac{R_3-R_2-R_1}{2}=\ln\left(\sqrt{1-\lambda_1 \lambda_2}\left|\frac{k_1-k_2}{k_1+k_2^*}\right| \right) \equiv -\Phi_2,
\label{phase}\eea
where $R_1=\ln(\frac{\kappa_{11}}{k_1+k_1^*})$, $R_2=\ln(\frac{\kappa_{22}}{k_2+k_2^*})$ and $R_3=\ln\left(\frac{|k_1-k_2|^2(\kappa_{11}\kappa_{22}-\kappa_{12}\kappa_{21})}{(k_1+k_1^*)|k_1+k_2^*|^2(k_2+k_2^*)}\right)$. On the other hand, the solitons appearing in the LW component undergo only elastic collision with a phase-shift (\ref{phase}) for all the choices of polarization parameter $\alpha_j^{(\ell)}$. The phase-shift experienced by the colliding solitons results in a change in the relative separation distance between the two solitons. This can be defined as the difference between the relative separation distances between the solitons before collision ($d_{12}^{-}=\frac{(R_3-R_1)k_{1R}-R_1 k_{2R}}{2k_{1R}k_{2R}}$) and after collision ($d_{12}^{+}=\frac{R_2 k_{1R}-(R_3-R_2)k_{2R}}{2k_{1R}k_{2R}}$) and its exact form is obtained as
\bea \Delta=d_{12}^{-}-d_{12}^{+} \equiv \frac{k_{1R}+k_{2R}}{k_{1R}k_{2R}}\Phi_1. \label{distance}\eea The soliton collision scenario and the bound solitons of system (\ref{model}) with $c_{\ell}=1$ and $\delta=1$ have been studied in detail by the present authors in their earlier works \cite{Kanna2009jpa,Sakkara2013epjst}. A detailed study on the dynamics of bright soliton bound state of the present system for arbitrary nonlinearities can also be carried out as done in Ref. \cite{Sakkara2013epjst}. Below we discuss the collision scenario of bright solitons for different choices of $c_{\ell}$.

\subsection*{Case (i): Positive nonlinearity coefficients ($c_{\ell}>0$)}
\begin{figure}[h]
\centering\includegraphics[width=0.933\linewidth]{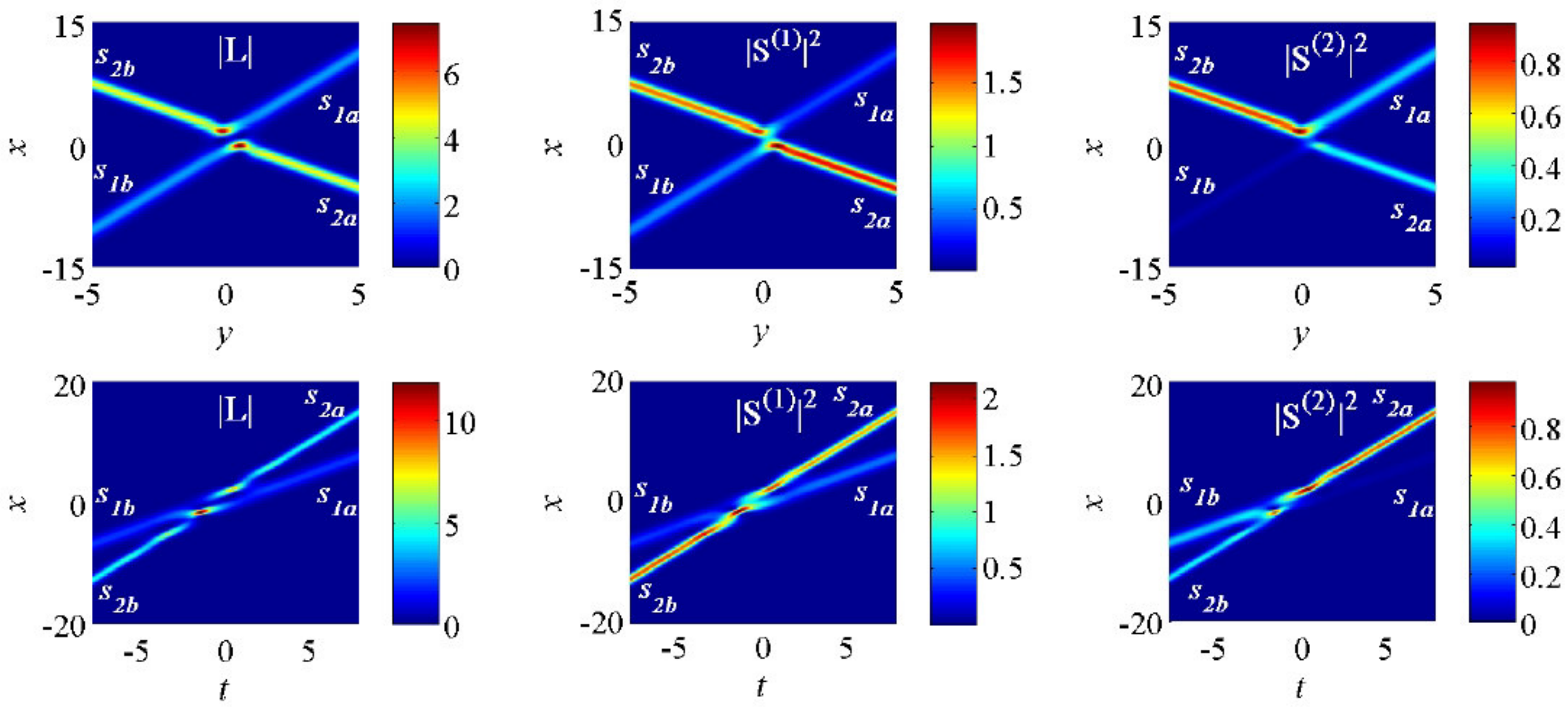}%2a.eps}\includegraphics[width=0.33\linewidth]{figure2b.eps}\includegraphics[width=0.33\linewidth]{figure2c.eps}\\
%\centering\includegraphics[width=0.33\linewidth]{figure2d.eps}\includegraphics[width=0.33\linewidth]{figure2e.eps}\includegraphics[width=0.33\linewidth]{figure2f.eps}
\caption{(Color online) Type-I energy sharing collision of two bright solitons in 2-LSRI system for positive nonlinearity coefficients ($c_1,c_2>0$). Head-on collision of solitons in the ($x-y$) plane at $t=1$ (top panels) and overtaking collision of solitons in the ($x-t$) plane at $y=1$ (bottom panels). The parameters are $c_1=2$, $c_2=1$, $k_1=1-1.5i$, $k_2=1.5-0.25i$, $\omega_1=-1-i$, $\omega_2=-2.5-0.5i$, $\alpha_1^{(1)}=2$, $\alpha_2^{(1)}=1.1$, $\alpha_1^{(2)}=0.6$, $\alpha_2^{(2)}=0.5$. In Figs. 3-8, $s_{jb}$ and $s_{ja}$ refer to the soliton $s_j,~j=1,2,$ before and after collision.}
\label{sc1fig1a}
\end{figure}

The energy sharing collision scenario of two bright solitons for this case in the ($x-y$) plane (($x-t$) plane) is shown in the top (bottom) panels of Fig. \ref{sc1fig1a}.  In the ($x-y$) plane, the amplitude of soliton $s_1$ is enhanced (suppressed) in the $S^{(1)}$ ($S^{(2)}$) component after collision, while the change in the amplitude of soliton $s_2$ is just opposite to that of $s_1$ in a given SW component. Both the colliding solitons experience a phase-shift as give by (\ref{phase}). The switching nature of soliton intensity (energy) during collision in the ($x-t$) plane is reversed as compared with the collision scenario in the ($x-y$) plane. But, the LW solitons emerge unaltered after collision only with a phase-shift in both the ($x-y$) and ($x-t$) planes. In the ($x-y$) plane, solitons can undergo both head-on and overtaking collisions, whereas in the ($x-t$) plane they are restricted to undergo only overtaking collisions since the condition for regular soliton requires $\frac{\omega_{1R}}{k_{1R}}$ and $\frac{\omega_{2R}}{k_{2R}}$ to be positive or negative simultaneously, which means that both the solitons should propagate in the same direction but with different speeds. Notice that a given soliton (either $s_1$ or $s_2$) experiences an opposite kind of energy switching in two SW components. In this collision process, the energy of solitons in individual SW (and LW) component and also the total energy of solitons among all the SW components are conserved. We refer to such a collision process as a type-I energy-sharing collision. Such a type-I energy-sharing collision has previously been observed in multicomponent Manakov type system \cite{Kanna-cnls} and also in multicomponent Yajima-Oikawa system \cite{Kanna2013pre}.

\begin{figure}[h]
\centering\includegraphics[width=0.933\linewidth]{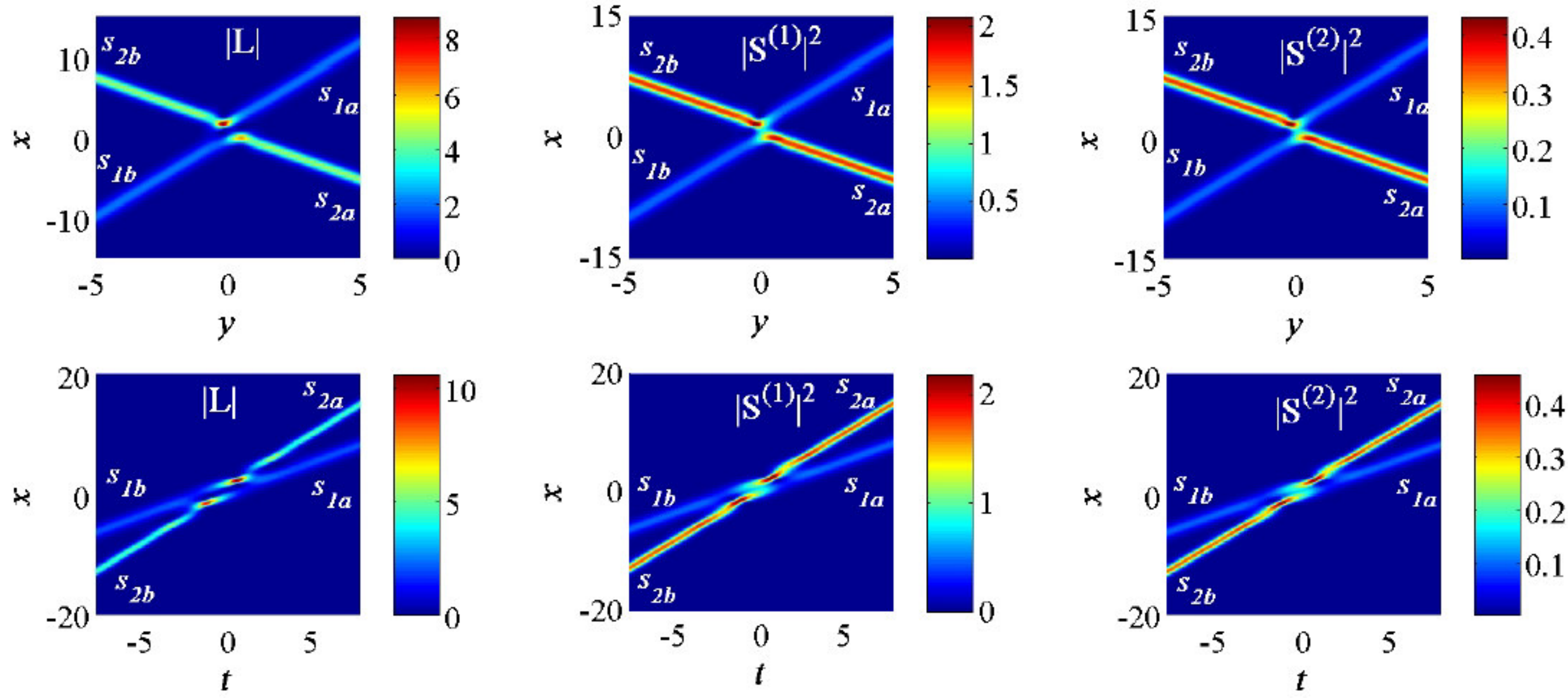}%3a.eps}\includegraphics[width=0.33\linewidth]{figure3b.eps}\includegraphics[width=0.33\linewidth]{figure3c.eps}\\
%\centering\includegraphics[width=0.33\linewidth]{figure3d.eps}\includegraphics[width=0.33\linewidth]{figure3e.eps}\includegraphics[width=0.33\linewidth]{figure3f.eps}
\caption{(Color online) Elastic collision of two bright solitons in 2-LSRI system for positive nonlinearity coefficients.}% Here the parameters are same to that of Fig. \ref{sc1fig1a} except for $\alpha_1^{(1)}=\alpha_2^{(1)}=1.7$ and $\alpha_1^{(2)}=\alpha_2^{(2)}=0.5$.}
\label{elasfig1}
\end{figure}

The elastic collision of solitons in the ($x-y$) as well as in the ($x-t$) planes are depicted in Fig. \ref{elasfig1} for the choices $\alpha_1^{(1)}=\alpha_2^{(1)}=1.7$ and $\alpha_1^{(2)}=\alpha_2^{(2)}=0.5$, by keeping the other parameters fixed as in Fig. \ref{sc1fig1a}. Here, the amplitudes of both solitons, $s_1$ and $s_2$ remain unaltered after collision in all the three components (2 SWs and 1 LW components). But they suffer a phase-shift after collision.

\subsection*{Case (ii): Negative nonlinearity coefficients ($c_{\ell}<0$)}
Bright solitons in 2-LSRI system with negative nonlinearity coefficients ($c_{\ell}<0$) admit similar type of collision behavior as that of the positive nonlinearity case ($c_{\ell}>0$). Here also the solitons undergo both head-on and overtaking collisions in the ($x-y$) plane and there occurs only overtaking collision in the ($x-t$) plane, involving energy-sharing nature of type-I or elastic nature of soliton amplitudes accompanied by a phase-shift. The only difference is that the signs of both $k_{jR}$ and $\omega_{jR}$ have to be fixed as positive for obtaining regular solitons (non-singular solution) and hence the direction of overtaking collision in the ($x-t$) plane gets reversed.

\subsection*{Case (iii): Mixed-type nonlinearity coefficients}
\begin{figure}[h]
\centering\includegraphics[width=0.933\linewidth]{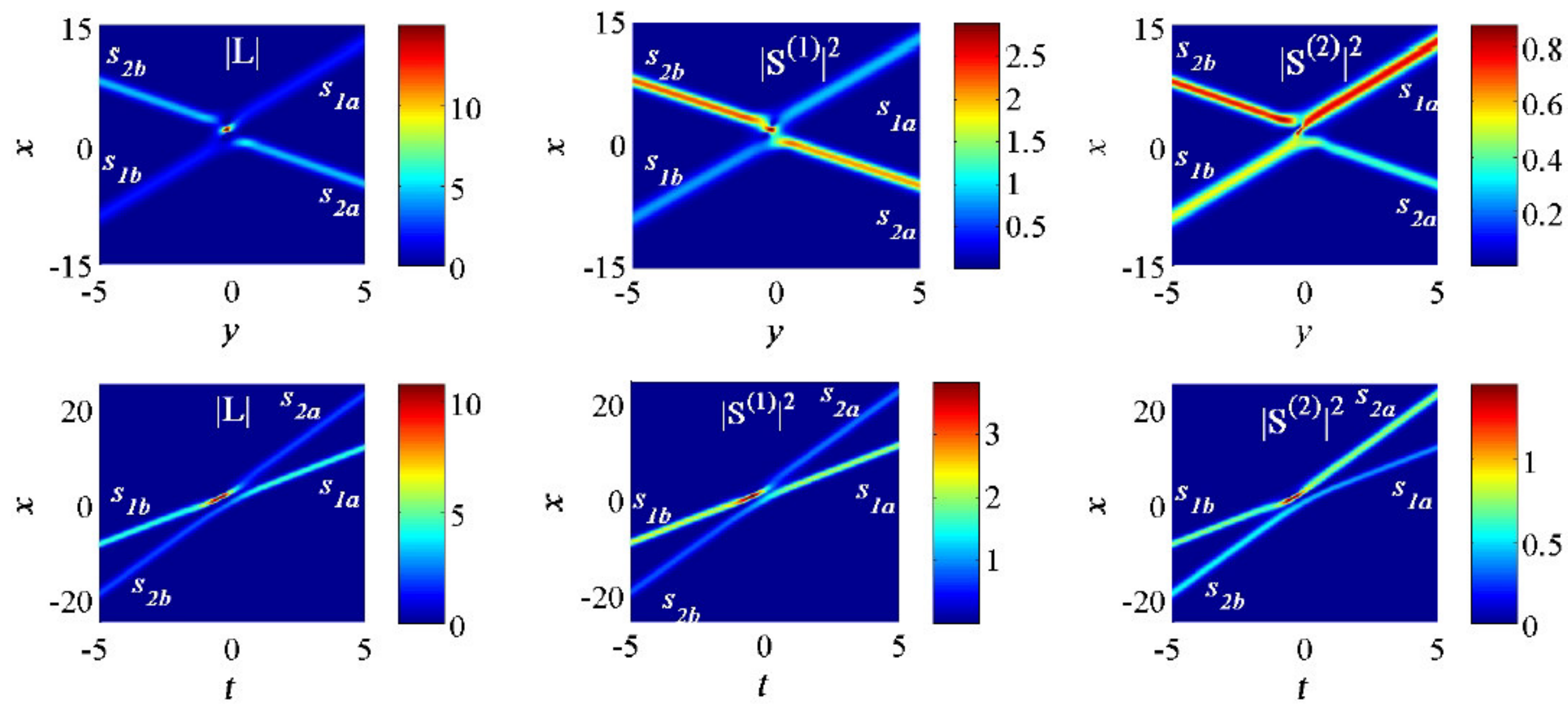}%4a.eps}\includegraphics[width=0.33\linewidth]{figure4b.eps}\includegraphics[width=0.33\linewidth]{figure4c.eps}\\
%\centering\includegraphics[width=0.33\linewidth]{figure4d.eps}\includegraphics[width=0.33\linewidth]{figure4e.eps}\includegraphics[width=0.33\linewidth]{figure4f.eps}
\caption{(Color online) Type-II energy sharing collision of two bright solitons of 2-LSRI system for mixed type nonlinearity coefficients. Soliton collision in the ($x-y$) plane (top panels) for $t=1$ and in the ($x-t$) plane (bottom panels) for $y=2$. Here the parameters are $c_1=2$, $c_2=-1$, $k_1=1-1.5i$, $k_2=1.5-0.25i$, $\omega_1=-1-i$, $\omega_2=-2.5-0.5i$, $\alpha_1^{(1)}=0.6$, $\alpha_2^{(1)}=0.7$, $\alpha_1^{(2)}=0.5$ and  $\alpha_2^{(2)}=0.3$. }
\label{sc2fig}
\end{figure}

For mixed type nonlinearity coefficients $c_{\ell}$ ($c_{\ell}>0$, $\ell=1,2,...,Q$ and $c_{\ell}<0$, $\ell=Q+1,Q+2,...,M$) the bright solitons display a different type of  energy sharing collision of solitons. We have shown such an energy sharing collision in Fig. \ref{sc2fig}.
It can be noticed that the amplitude of soliton $s_1$ gets suppressed while that of soliton $s_2$ gets enhanced after collision in both SW components. This  same type of energy switching of solitons among all SW components results due to a special kind of soliton energy conservation. Here the energy in individual component and also the difference of energy between the SW components are conserved, rather than the conservation of total energy among the components as in the cases (i) and (ii). We refer to such collision process as type-II energy sharing collision. This type-II energy-sharing collision of solitons has been observed in a particular one-dimensional integrable CNLS system \cite{Kanna-cnls} and also recently in multicomponent Yajima-Oikawa system \cite{Kanna2013pre} with focusing-defocusing (mixed) type of nonlinearities. To the best of our knowledge, for the first time we report such type-II energy sharing collision in higher-dimensional system. The solitons in the LW component emerge elastically after collision with a phase-shift given by Eq. (\ref{phase}). Thus, for the mixed type nonlinearity coefficients the nature of switching of intensities (energy) for a given soliton is same in all the $M$ number of SW components in the $M$-LSRI system (1), while in a given component the colliding solitons $s_1$ and $s_2$ experience an opposite kind of energy switching. This enables the present system (1) to achieve amplification of a particular soliton in all the $M$ components after collision with other soliton. Typical type-II energy-sharing collisions in the ($x-y$) plane (($x-t$) plane) are shown in the top (bottom) panels of Fig. \ref{sc2fig}.

We have also noticed that multi-soliton collision, collision involving more than two solitons, which takes place in a pair-wise manner. So, based on the above two-soliton collision scenario one can easily investigate the features of a multi-soliton collision of system (\ref{model}).

\section{Resonant Solitons}\label{secreso}
The general $n$-soliton solution (\ref{nsol}) features special localized structures, namely resonant solitons, in addition to the standard interacting solitons. These resonant solitons can be achieved by appropriately choosing the soliton parameters such that the phase-shifts due to collision become infinity, i.e. $|\Phi_1|=|\Phi_2|=\infty$ [see Eq. (\ref{phase})]. Thus the resonant soliton is a localized wave that appears in the interaction regime and can be viewed as an intermediate state during soliton interaction. The reason for the existence of such a long-living intermediate state is that as the phase-shifts of colliding solitons approach infinity, the span of the interaction regime also extends to infinity before the solitons get well separated. This is due to the fact that the change in the relative separation between the solitons [see Eq. (\ref{distance})] becomes infinity. In such a case, it is possible to achieve large amplitude localized wave structures (the resonant soliton) for infinite distance/time.
\begin{figure}[h]
\centering\includegraphics[width=0.633\linewidth]{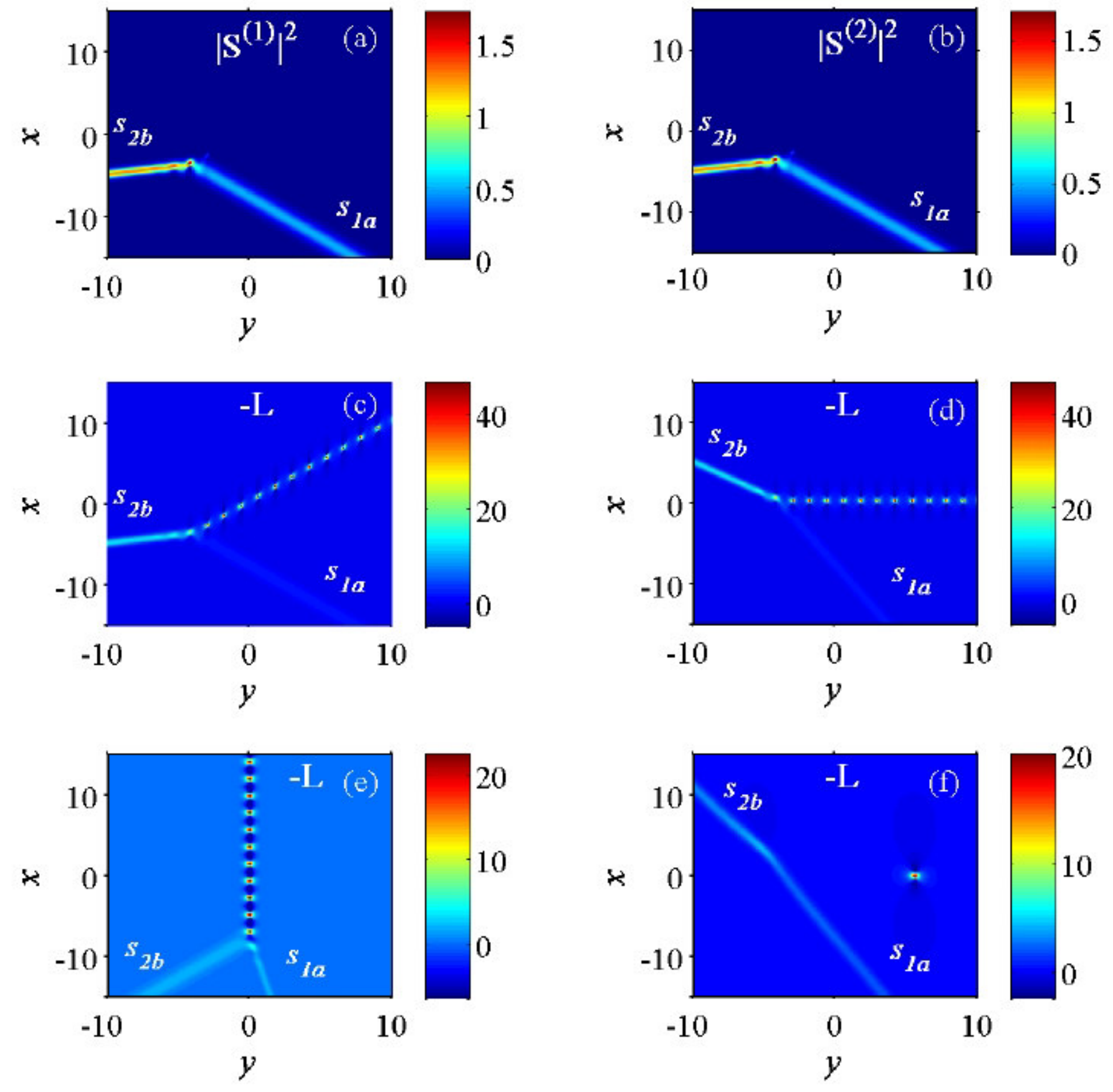}%5a.eps}\includegraphics[width=0.33\linewidth]{figure5b.eps}\\
%\centering\includegraphics[width=0.33\linewidth]{figure5c.eps}\includegraphics[width=0.33\linewidth]{figure5d.eps}\\
%\centering\includegraphics[width=0.33\linewidth]{figure5e.eps}\includegraphics[width=0.33\linewidth]{figure5f.eps}
\caption{(Color online) Resonant solitons (breathers) in the 2-LSRI system for $\omega_1=\omega_2=-2$,  $\alpha_1^{(1)}=\alpha_1^{(2)}=\alpha_2^{(1)}=\alpha_2^{(2)}=1$ and $c_1=c_2=2$. Zero-amplitude resonant soliton in the SW components (a--b) and a general breather-like structure in the LW component (c) for $k_1=1-0.5$ and $k_2=2.5-0.5i$, resonant-breather localized in '$x$' for $k_1=1$ and $k_2=2.5$ (d), resonant-breather localized in `$y$' for $k_1=1+1.5i$, $k_2=1-1.5i$ (e) and a localized structure in both $x$ and $y$ coordinate co-existing with solitons for $k_1=1+0.05i$, $k_2=1.25+0.05i$ (f).}
\label{reso}
\end{figure}

The present $M$-LSRI system (\ref{model}) supports RS for particular choice of soliton parameters, $\omega_j$, $j=1,2$. We find from Eq. (\ref{phase}) for the phase-shifts that the RS is possible when $1-\lambda_1 \lambda_2=0$, which results in infinite magnitude for $\Phi_j$. This can be obtained by setting $\omega_1=\omega_2$ and for this choice the two colliding solitons approach each other only asymptotically and forms a RS which is very distinct from the standard interacting solitons. This RS in the LW component exhibits localized periodic structures in the ($x-y$) plane, similar to the breathers on a zero background \cite{zerobreather}, and we refer to these structures as resonant-breathers (RBs) in analogy with the standard breather in soliton theory [see Figs. \ref{reso}c--\ref{reso}e]. The localization of these RBs can be either in `$x$ or in `$y$' coordinate depending on the values of $k_{jR}$ and $k_{jI}$, $j=1,2$.

We have shown such resonant solitons (breathers) in Fig. \ref{reso} for the positive nonlinearity coefficients ($c_{\ell}>0$). Here the two colliding solitons form a zero-amplitude resonant state in the interaction regime of two SW components. However, in the LW component, the resonant soliton looks similar to the breather on a zero background with periodic oscillations in its amplitude. An interesting physical process to be noticed in the formation of resonant soliton in the interaction regime is that the energy of SW components completely disappears and reappears in the LW component, thereby resulting in a resonant soliton with large amplitude localized structure with periodic oscillations in that LW component. This is a consequence of total energy conservation of the system (\ref{model}).

When $k_{1R}\neq k_{2R}$ and $k_{1I}=k_{2I}$, in addition to $\omega_1=\omega_2$, we get the resonant breather (Fig. \ref{reso}c) and this will be localized along $x$ coordinate for $k_{1I}=k_{2I}=0$ (Fig. \ref{reso}d). For the choice $k_{1R}= k_{2R}$ and $k_{1I} \neq k_{2I}$, the resonant breather will be localized in the $y$ coordinate (Fig. \ref{reso}e). It is also possible to localize the resonant-breather both in `$x$' and `$y$' coordinates by tuning the $k_{jR}$ and $k_{jI}$ parameters suitably. Particularly, for $k_{1R} \simeq k_{2R}$  and $k_{1I}= k_{2I}$ (or $k_{1R} = k_{2R}$  and $k_{1I} \simeq k_{2I}$) we have non-trivial resonant-breather co-existing with solitons as shown in Fig. \ref{reso}f. This localized structure looks akin to a rogue wave but in a zero-background co-existing with regular solitons. This localized structure is a special feature of the (2+1)D LSRI system (\ref{model}) and has been predicted here for the first time to the best of our knowledge. It will be relevant to point out that the co-existence of rogue wave in the non-zero background with dark-bright soliton has been observed in a two-component vector NLS system with focusing nonlinearity \cite{DegasPRL}.

\begin{table}[h]
\caption{Nature of resonant solitons (breathers) of system (1)}
\scriptsize
\label{table2}
\begin{tabular}{| c| p{1.5cm}| p{2cm}| c| p{6cm}| p{1cm} |}
  \hline \hline
  Case  & Choice of $k_{jR}$ & Choice of $k_{jI}$ & Choice of $\omega_j$ & RS/RB in ($x-y$) plane of the LW component & Figure No.\\ \hline \hline
  (i) & $k_{1R} \neq k_{2R}$ & $k_{1I} = k_{2I} \neq 0$ & $\omega_1 = \omega_2$ & RB:- infinite length breather  & \ref{reso}c \\\hline
  (ii) & $k_{1R} \neq k_{2R}$ & $k_{1I} = k_{2I}=0$ & $\omega_1=\omega_2$ & RB:- breather localized in $x$  & \ref{reso}d \\\hline
  (iii) & $k_{1R}=k_{2R}$ & ~~~$k_{1I} \neq k_{2I}$ & $\omega_1=\omega_2$ & RB:- breather localized in $y$  & \ref{reso}e\\ \hline
  (iv) & $k_{1R} \simeq k_{2R}$ $k_{1R} = k_{2R}$ & $k_{1I} = k_{2I}$ \newline $k_{1I} \simeq k_{2I}$ & $\omega_1 = \omega_2$ & RB:- single excited structure localized in both `$x$' and `$y$' (can be viewed as rogue wave in zero background co-existing with solitons)  & \ref{reso}f \\\hline
  (v) & $k_{1R} \neq k_{2R}$ & $k_{1I} = k_{2I} \neq 0$ & $\omega_1 \simeq \omega_2$ & QRB:- breather of finite span & \ref{near-reso}a \\\hline
  (vi) & $k_{1R} \neq k_{2R}$ & $k_{1I} = k_{2I}=0$ & $\omega_1 \simeq \omega_2$ & QRB:- breather along $x$ but of finite duration & \ref{near-reso}b \\\hline
  (vii) & $k_{1R} = k_{2R}$ & ~~~$k_{1I} \neq k_{2I}$ & $\omega_1 \simeq \omega_2$ & QRB:- breather along $y$ but of finite duration & \ref{near-reso}c \\\hline
  \hline
\end{tabular}
\end{table}

\begin{figure}[h]
\centering\includegraphics[width=0.65\linewidth]{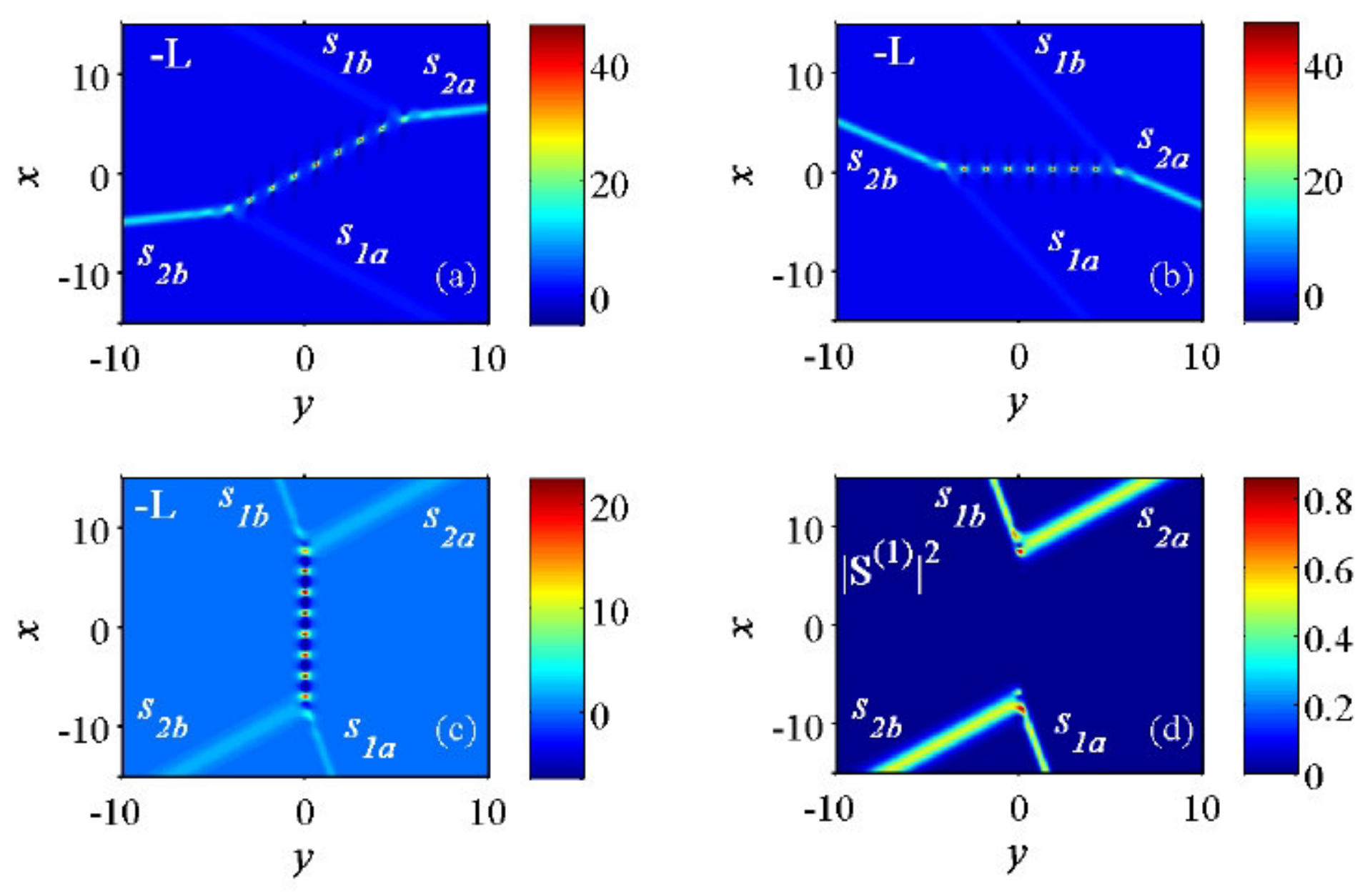}%6a.eps}\includegraphics[width=0.33\linewidth]{figure6b.eps}\\
%\centering\includegraphics[width=0.33\linewidth]{figure6c.eps}\includegraphics[width=0.33\linewidth]{figure6d.eps}
\caption{(Color online) Quasi-resonant breathers in the 2-LSRI system for $\omega_1=-2$, $\omega_2=-1.9999999$, $\alpha_1^{(1)}=\alpha_1^{(2)}=\alpha_2^{(1)}=\alpha_2^{(2)}=1$ and $c_1=c_2=2$. QRB for $k_1=1-0.5$ and $k_2=2.5-0.5i$ (a), the QRB localized in `$x$' for $k_1=1$ and $k_2=2.5$ (b), the QRB localized in `$y$' (c) and the zero-amplitude resonant soliton in the SW component (d) for $k_1=1+1.5i$, $k_2=1-1.5i$.}
\label{near-reso}
\end{figure}

Noticeably, for $\omega_1 \simeq \omega_2$, the two colliding solitons form a quasi-resonant breather (QRB) in which the intermediate interaction regime exists only for a finite duration and after that it splits into two individual solitons. Such quasi-resonant breathers are shown in Fig. \ref{near-reso}. Here we can obtain the resonant breather for a finite length and can be localized either in $x$ or $y$. But we can not obtain the rogue-wave-like structure co-existing with solitons for this quasi-resonant choice. The main difference between Figs. \ref{near-reso} and \ref{reso} is that the span of the resonant breather is finite in the former while it extends up to infinity in the latter. We have summarized the above details on resonant solitons (breathers) in Table \ref{table2}.

\begin{figure}[h]
\centering\includegraphics[width=0.633\linewidth]{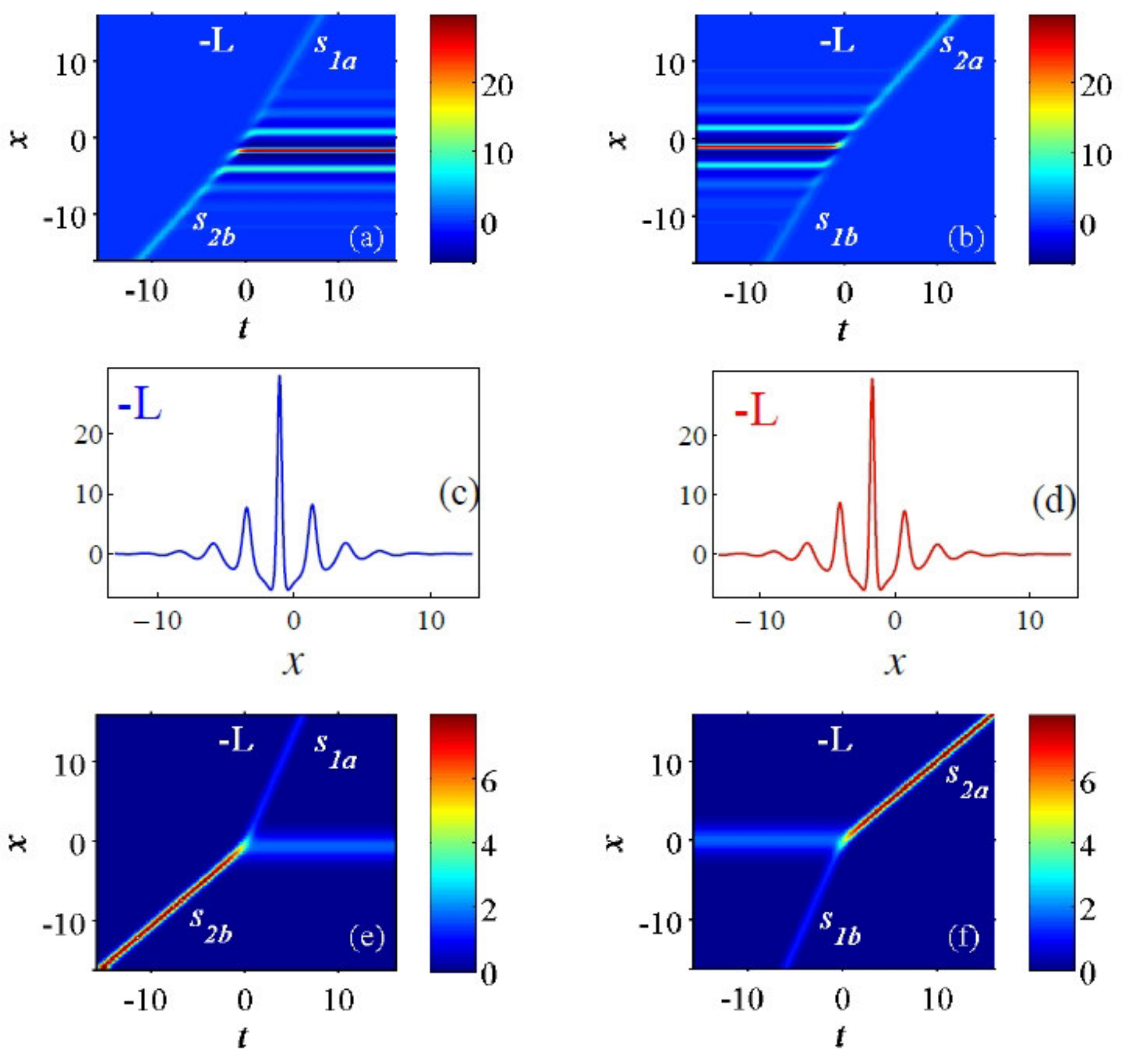}%7a.eps}\includegraphics[width=0.33\linewidth]{figure7b.eps}\\
%\centering\includegraphics[width=0.25\linewidth]{figure7e.eps}~~~~~~~~~\includegraphics[width=0.25\linewidth]{figure7f.eps}\\
%\centering\includegraphics[width=0.33\linewidth]{figure7c.eps}\includegraphics[width=0.33\linewidth]{figure7d.eps}
\caption{(Color online) Resonant solitons in the ($x-t$) plane of the LW component of 2-LSRI system with oscillating side-bands (a--d) and without oscillations (e--f) at $y=-1$. The parameters are $c_1=c_2=2$ with $k_1=1+0.5i$, $k_2=1.5-2i$, $\omega_1=\omega_2=-2$ (a), $k_1=-1+0.5i$, $k_2=-1.5-2i$, $\omega_1=\omega_2=2$ (b), $k_1=0.75-1.5i$, $k_2=2-1.5i$, $\omega_1=\omega_2=-2$ (e), and $k_1=-0.75-1.5i$, $k_2=-2-1.5i$, $\omega_1=\omega_2=2$ (f). The 2D plots (c) and (d) clearly show the sideband oscillations corresponding to figures (a) and (b), respectively at $t=12$ and $t=-12$.}
\label{reso-xt1}
\end{figure}

One can also demonstrate the existence of resonant solitons in the ($x-t$) plane similar to that of the ($x-y$) plane. Interestingly, in the ($x-t$) plane, the RS exhibits constant amplitude wave structure with oscillating side-bands (see Figs. \ref{reso-xt1}a--\ref{reso-xt1}d) and the number of sidebands depends upon the absolute difference between $k_{1I}$ and $k_{2I}$ (that is, $|k_{1I}-k_{2I}|$). When $|k_{1I}-k_{2I}|\rightarrow 0$, the sidebands disappear and we get the RS with a single peak wave structure (see Figs. \ref{reso-xt1}e--\ref{reso-xt1}f) and the resulting collision scenario depicted in Figs. \ref{reso-xt1}e and \ref{reso-xt1}f look like the process of soliton fission and fusion respectively. Unlike in the case of resonant solitons in the ($x-y$) plane, here in the ($x-t$) plane the resonant soliton can only be localized along the $x$ direction.

A similar analysis on the dynamics of resonant solitons for other choices of arbitrary nonlinearity coefficients can be performed by extending the above type of investigation. One can also perform a detailed analysis on the dynamics of multi-soliton resonance in view of the above discussion and can explore different (web-like) structures.

\section{Dark soliton solutions}\label{secdark}
For completeness, in this section we obtain dark-soliton solutions of the $M$-LSRI system (1) using the Hirota's bilinearization method \cite{Kiv1993ol,Kiv1997pre,Ohtadark}. For this purpose, let us consider the constant parameter ($\lambda$) appearing in the bilinear equations (2) to be non-zero. This will lead us to obtain the dark solitons in an asymptotically non-vanishing limit, i.e. $|S^{(\ell)}|\rightarrow \mbox{constant}$ and $|L|\rightarrow 0$ when $x,y,t\rightarrow \pm \infty$. To construct the $n$-dark soliton solution, one has to consider the form of power series expansion for $g^{(\ell)}$ and $f$ as $g^{(\ell)}=g_0^{(\ell)}\left(1+\sum_{j=1}^n \chi^{2j} g_{2j}^{(\ell)}\right)$, $\ell=1,2,...,M$, and $f=1+\sum_{j=1}^n \chi^{2j} f_{2j}$. Considering the length of the paper, we explicitly construct the one- and two- soliton solutions in this section and one can extend the present algorithm straightforwardly to obtain the $n$-dark soliton solution for arbitrary $n$.

\subsection{Dark one-soliton solution}
To construct the dark one-soliton solution ($n=1$), we terminate the power series expansion as $g^{(\ell)}=g_0^{(\ell)}(1+\chi^2 g_2^{(\ell)})$, $\ell=1,2,...,M$, and $f=1+\chi^2 f_2$. Then by recursively solving the resulting set of bilinear equations after substituting the above power series into Eq. (2), we obtain the one-dark soliton solution as
\bes\bea
S^{(\ell)}&=&\frac{g^{(\ell)}}{f} = \tau_{\ell}\left(\frac{1+\psi_1^{(\ell)}e^{\eta_1}}{1+ e^{\eta_1}}\right)e^{i\xi_{\ell}}, \quad \ell=1,2,...,M,\\
L&=&-2\frac{\partial^2}{\partial x^2}\left[\ln(1+ e^{\eta_1})\right],
\eea\label{1ds}\ees
where $\eta_1=k_1 x+p_1 y+\omega_1 t$, $\xi_{\ell}=a_{\ell} x+b_{\ell} y+\sigma_{\ell} t$, $\lambda=\sum_{\ell=1}^M c_{\ell}|\tau_{\ell}|^2$, and $\psi_1^{(\ell)}=\frac{2a_{\ell}k_1-\delta p_1-\omega_1+ik_1^2}{2a_{\ell}k_1-\delta p_1-\omega_1-ik_1^2}$. The dark one-soliton solution (\ref{1ds}) is characterized by ($4M+3$) real parameters ($c_{\ell}$, $a_{\ell}$, $b_{\ell}$, $\sigma_{\ell}$, $k_1$, $p_1$ and $\omega_1$) and $M$ complex parameters ($\tau_{\ell}$) with ($M+1$) conditions, $\sigma_{\ell}=a_{\ell}^2-\delta b_{\ell}$, $\ell=1,2,...,M$, and $\frac{4 k_1^3}{\omega_1}\ds\sum_{\ell=1}^M \frac{c_{\ell} |\tau_{\ell}|^2}{(2a_{\ell}k_1-\delta p_1-\omega_1)^2+k_1^4}= 1$. The above dark one-soliton solution (\ref{1ds}) can be rewritten as
\bes\bea
S^{(\ell)}&=& \frac{\tau_{\ell}}{2} \left[(1+\psi_1^{(\ell)})-(1-\psi_1^{(\ell)})\tanh(\eta_1/2)\right] e^{i\xi_{\ell}}, \quad \ell=1,2,...,M,\\
L&=& -\frac{k_1^2}{2}~\mbox{sech}^2(\eta_1/2).
\eea \label{1dsa} \ees

\begin{figure}[h]
\centering\includegraphics[width=0.933\linewidth]{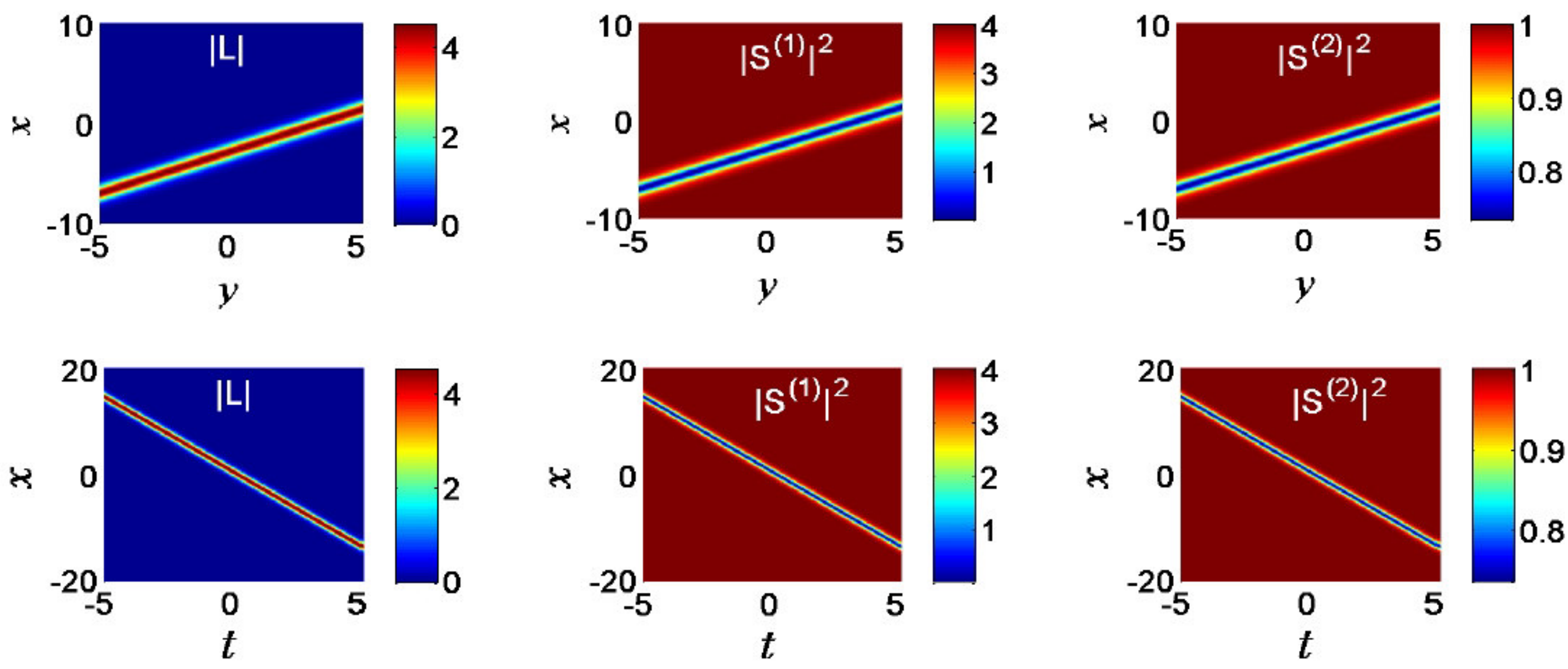}%8a}\includegraphics[width=0.33\linewidth]{figure8b}\includegraphics[width=0.33\linewidth]{figure8c}\\
%\centering\includegraphics[width=0.33\linewidth]{figure8d}\includegraphics[width=0.33\linewidth]{figure8e}\includegraphics[width=0.33\linewidth]{figure8f}
\caption{Propagation of bright ($|L|$), dark ($|S^{(1)}|^2$) and gray ($|S^{(2)}|^2$) solitons in the 2-LSRI system in the ($x-y$) plane for $t=1$ (top panels) and in the ($x-t$) plane for $y=0.5$ (bottom panels).}
\label{dsfig1s}
\end{figure}

Also, the intensity of the dark-soliton in SW components can be written in a compact form as $|S^{(\ell)}|^2= |\tau_{\ell}|^2 \left[1-A_{\ell}~ \mbox{sech}^2(\eta_1/2)\right]$, where $A_{\ell}=\frac{k_1^4}{(2 a_{\ell} k_1-\delta p_1-\omega_1)^2+k_1^4}$, $\ell=1,2,...,M$. Here the degree of darkness of the soliton in $\ell$-th SW component can be determined by the quantity $A_{\ell}$ and background intensity $|\tau_{\ell}|^2$. One can obtain a dark or gray soliton in the SW components by tuning $A_{\ell}$ appropriately, i.e. for $A_{\ell}=1$ ($A_{\ell}<1$) we get dark (gray) soliton in the SW components. Also, we can profitably control the dark soliton profile by tuning the nonlinearity coefficients $c_{\ell}$ in addition to the other soliton parameters, as it explicitly appears in the condition for those soliton parameters. But, the LW component always results in bright soliton with amplitude $\frac{-k_1^2}{2}$ and it does not depend on the other soliton parameters. However, all the solitons (bright in LW and dark in SW) travel with velocity $-\frac{p_1}{k_1}$ in the ($x-y$) plane and $-\frac{\omega_1}{k_1}$ in the ($x-t$) plane. The propagation of bright (dark) soliton appearing in the LW (SW) component(s) of 2-LSRI system is shown in Fig. \ref{dsfig1s} for the choice of parameters $\delta=1$, $c_1=1.5$, $c_2=1$, $k_1=3$, $p_1=-2.5$, $a_1=1$, $a_2=-1.52$, $b_1=1.1$, $b_2=1.3$, $\tau_1=2$ and $\tau_2=1$.

\subsection{Dark two-soliton solution and their collision}
We obtain the following dark two-soliton solution ($n=2$) by restricting the power series expansions for $g^{(\ell)}$ and $f$ as $g^{(\ell)}=g_0^{(\ell)}(1+\chi^2 g_2^{(\ell)}+\chi^4 g_4^{(\ell)})$, $\ell=1,2,...,M$, and $f=1+\chi^2 f_2+\chi^4 f_4$ and by solving the resultant bilinear equations arising at various powers of $\chi$:
\bes\bea
S^{(\ell)}&=& \tau_{\ell} \left(\frac{1+\psi_1^{(\ell)}e^{\eta_1}+\psi_2^{(\ell)}e^{\eta_2}+\psi_1^{(\ell)}\psi_2^{(\ell)}\Omega e^{\eta_1+\eta_2}}{1+ e^{\eta_1}+ e^{\eta_2}+\Omega  e^{\eta_1+\eta_2}}\right)e^{i\xi_{\ell}}, \quad \ell=1,2,...,M,\\
L&=&-2\frac{\partial^2}{\partial x^2}\left[\ln(1+ e^{\eta_1}+ e^{\eta_2}+\Omega  e^{\eta_1+\eta_2})\right],
\eea \label{2ds} \ees
where $\eta_j=k_j x+p_j y+\omega_j t$, $\xi_{\ell}=a_{\ell} x+b_{\ell} y+\sigma_{\ell} t$, $\lambda=\sum_{\ell=1}^M c_{\ell}|\tau_{\ell}|^2$, $\psi_j^{(\ell)}=\frac{2a_{\ell}k_j-\delta p_j-\omega_j+ik_j^2}{2a_{\ell}k_j-\delta p_j-\omega_j-ik_j^2}$, $j=1,2$, $\ell=1,2,...,M$, and $\Omega =\frac{k_1^2 k_2^2(k_1-k_2)^2+(k_1(\delta p_2+\omega_2)-k_2(\delta p_1+\omega_1))^2}{k_1^2 k_2^2(k_1+k_2)^2+(k_1(\delta p_2+\omega_2)-k_2(\delta p_1+\omega_1))^2}$. The above dark two-soliton solution is characterized by ($4M+6$) real parameters ($c_{\ell}$, $a_{\ell}$, $b_{\ell}$, $\sigma_{\ell}$, $k_j$, $p_j$ and $\omega_j$, $\ell=1,2,...,M,$ $j=1,2$) and $M$ complex parameters ($\tau_{\ell}$), with $(M+2)$ conditions $\sigma_{\ell}=a_{\ell}^2-\delta b_{\ell}$, $\ell=1,2,...,M$, and $\frac{2}{\omega_j k_j} \ds\sum_{\ell=1}^M c_{\ell}|\tau_{\ell}|^2 \left(1-Re[\psi_j^{(\ell)}]\right)=1$, $j=1,2$. As mentioned in the dark one-soliton solution, here also the velocities and amplitude (depth) of bright (dark) solitons can be controlled by tuning these arbitrary parameters.
\begin{figure}[h]
\centering\includegraphics[width=0.933\linewidth]{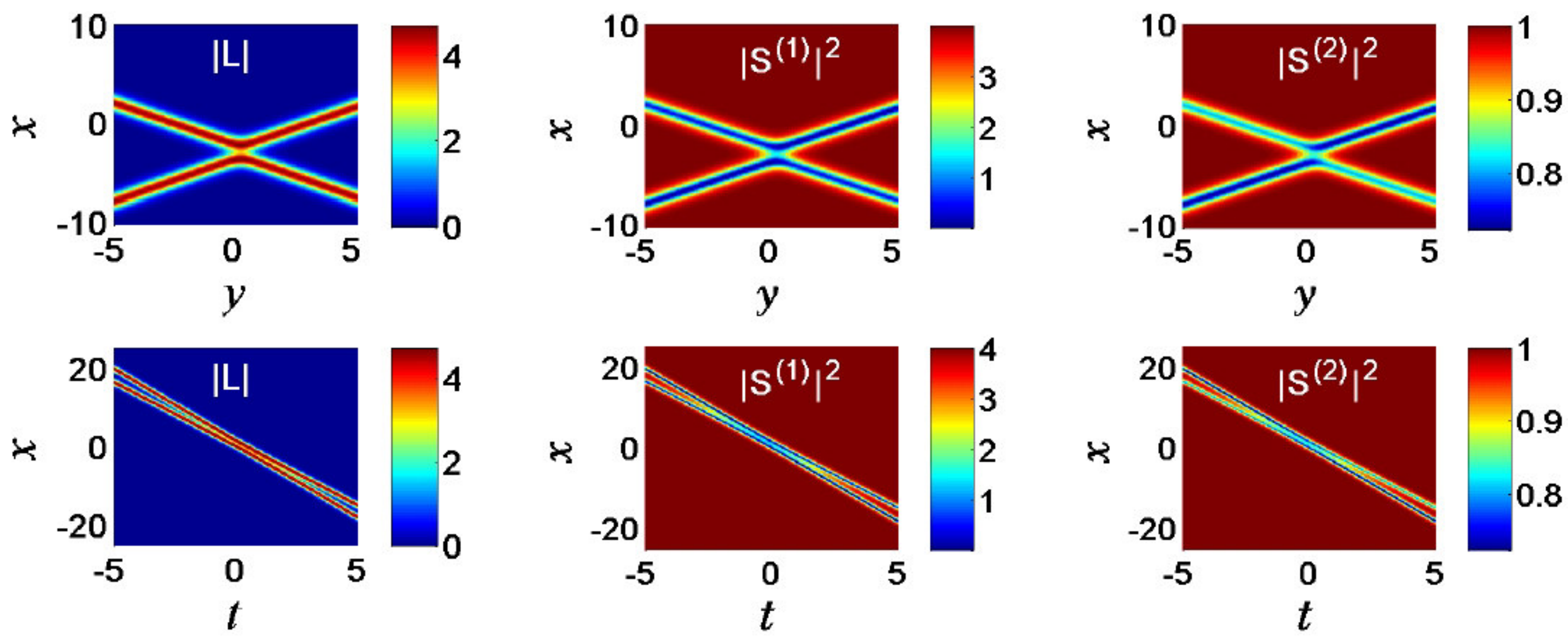}%9a}\includegraphics[width=0.33\linewidth]{figure9b}\includegraphics[width=0.33\linewidth]{figure9c}\\
%\centering\includegraphics[width=0.33\linewidth]{figure9d}\includegraphics[width=0.33\linewidth]{figure9e}\includegraphics[width=0.33\linewidth]{figure9f}
\caption{Elastic collision of bright-bright, dark-dark and dark-gray solitons, respectively in the $L$, $S^{(1)}$ and $S^{(2)}$ components of 2-LSRI system. Soliton collisions in the ($x-y$) plane for $t=1$ and in the ($x-t$) plane for $y=0.5$ with other parameters as $\delta=1$, $c_1=2$, $c_2=1$, $k_1=3$, $k_2=3,~p_1=-2.5$ and $p_2=2.5$, $a_1=1.5$, $a_2=-1$, $b_1=1.5$, $b_2=2$, $\tau_1=2$ and $\tau_2=1$.}
\label{fig2ds}
\end{figure}

By performing an asymptotic analysis of the dark soliton solution (\ref{2ds}), we found that the dark solitons in the SW components and the bright solitons appearing in the LW component undergo only elastic collisions for all choices of soliton parameters and they do not display any energy-sharing collision as in the case of bright short-wave solitons. The amplitude (depth) of bright (dark) solitons can be tuned by altering the soliton parameters and hence one can obtain either dark or gray soliton collision in the SW components. However, irrespective of the nature of soliton profile (bright in LW or dark (gray) in SW), they emerge unaffected after collisions. We have shown such elastic collision of two dark solitons of 2-LSRI system in Fig. \ref{fig2ds}.

In a straightforward manner, by following the above algorithm, one can construct the dark multi-soliton solutions and investigate their underlying dynamics.

\section{Summary and Conclusions}\label{conclusion}
We have considered a higher dimensional multicomponent long-wave--short-wave resonance interaction ((2+1)D $M$-LSRI) equations with arbitrary nonlinearity coefficients, describing the nonlinear resonant interaction of multiple short waves with a long-wave. By performing the Painlev\'e analysis we show that the general $M$-LSRI system (1) is integrable for arbitrary nonlinearity coefficients. Then we have obtained the bright multi-soliton solutions of $M$-LSRI system by using the Hirota's bilinearization method and presented the $n$-soliton solution in the form of Gram determinant. The significance of the arbitrary nonlinearity coefficients appearing in the solution has been examined by a detailed analysis of the propagation and collision dynamics of bright solitons. Particularly, we notice that the arbitrariness of the nonlinearity coefficients ($c_{\ell}$) results in a wider range of parameters for which the $M$-LSRI system supports regular solitons (non-singular solutions).

The higher dimensional bright solitons of $M$-LSRI system (1) have been classified into three cases based on the choices of signs of $c_{\ell}$, namely positive, negative and mixed signs. The dynamics of bright solitons for these three cases are explored in detail and has been summarized in the Table \ref{table1}. The change in the higher dimensional coefficient ($\delta$) does not alter the dynamical behavior of solitons but this alters the soliton velocity in the ($x-y$) plane and shifts the position of solitons in the ($x-t$) plane. We have shown that the analytical results are well in agreement with the numerical simulation.

Our study shows that the bright solitons undergo two kinds of energy sharing collision processes corresponding to $c_{\ell}>0$ ($c_{\ell}<0$) and mixed-signs for $c_{\ell}$ parameters and are referred to, respectively, as type-I and type-II energy-sharing collisions. Among them, the former conserves the total energy of all SW components while in the latter case the difference in the energy of SW components is conserved in addition to the conservation of energy in individual components. In addition to this, the SW solitons (for particular choice of $\alpha_j^{(\ell)}$ parameters) and LW solitons undergo standard elastic collision.

Finally, we have investigated the resonant soliton in the LSRI system (\ref{model}), which requires the phase-shifts of the colliding solitons to become infinity for its existence. We have shown that indeed it is possible to achieve infinite phase-shifts due to the higher dimensional nature of the present $M$-LSRI system. The resonant soliton exhibits a stable profile with large amplitude in the ($x-t$) plane and interestingly they form localized periodic structures with maximum amplitude similar to the breathers on a zero background in the ($x-y$) plane. By tuning the $k_j$, $j=1,2$, parameters we have demonstrated the appearance of a special localized structure (similar to rogue wave but in a zero-background) coexisting with the soliton. It is of future interest to investigate the link between the formation of rogue waves and multi-soliton resonances.

We have also obtained the dark one- and two- soliton solutions of $M$-LSRI system (\ref{model}) by using Hirota's direct method and briefly studied their propagation and collision dynamics. Through our analysis, we found that the depth (darkness) of the dark soliton can be controlled by tuning the soliton parameters. Also, the dark (gray) solitons always undergo elastic collision.
This study will find multifaceted applications, particularly in the context of optical computing \cite{steig}, nonlinear optics \cite{Ohta2007jpa,Kanna2013pre,Kanna-cnls,Kanna-ccnls,DegasPRL}, water wave theory \cite{Onorato-prl,Shukla-prl} and also in multicomponent Bose-Einstein condensates \cite{lsri-bec}.

\section*{ACKNOWLEDGMENTS}
The authors thank Dr. P. Muruganandam, School of Physics, Bharathidasan University, Tiruchirappalli-620 024, India, for his help in performing the numerical simulations. K.S. is grateful for the support of the Council of Scientific and Industrial Research, Government of India, with Senior Research Fellowship. The work of T.K. is supported by the Department of Science and Technology, Government of India, in the form of a major research project. M.V. acknowledges the financial support from UGC-Dr. D. S. Kothari post-doctoral fellowship scheme. The work of M.L. is supported by a DST-IRPHA project. M.L. is also supported by a DST Ramanna Fellowship project and a DAE Raja Ramanna Fellowship.

\appendix
\numberwithin{equation}{section}
\section{Painlev\'e analysis of (2+1)D $M$-LSRI system (1)}\label{secpain}
In this appendix, we present the results of the Painlev\'e singularity structure analysis \cite{Weiss,painml} of the (2+1)D $M$-LSRI system (\ref{model}). First, we rewrite the model equation (\ref{model}) as below by introducing a set of arbitrary real functions $u^{(\ell)}$, $v^{(\ell)}$ and $w$.
\bes\bea
&& i(u_t^{(\ell)}+\delta^{(\ell)} u_y^{(\ell)}) -  u^{(\ell)}_{xx}+  w u^{(\ell)}=0, \quad \ell=1,2,3,...,M,\\
&& -i(v_t^{(\ell)}+\delta^{(\ell)} v_y^{(\ell)}) -  v_{xx}^{(\ell)}+  w v^{(\ell)}=0,\quad \ell=1,2,3,...,M,\\
&& w_{t}=2\sum_{\ell=1}^M (c_{\ell} u^{(\ell)}v^{(\ell)})_x,
\eea\label{pa2}\ees
where $u^{(\ell)}=S^{(\ell)},~v^{(\ell)}=S^{(\ell)*}$ and $w=L$. The Painlev\'e analysis is carried out by expressing the dependent variables in terms of the following Laurent expansion in the neighbourhood of the non-characteristic manifold $\phi(x,y,t)$, with non-vanishing derivatives, $\phi_x,~\phi_y$ and $\phi_t$.
\bea
(u^{(\ell)},v^{(\ell)},w)=\left(\sum_{j=0}^N u_j^{(\ell)} \phi^{j+\alpha_{\ell}},\sum_{j=0}^N v_j^{(\ell)} \phi^{j+\beta_{\ell}},\sum_{j=0}^N w_j \phi^{j+\gamma}\right),\quad \ell=1,2,3,...,M. \label{ls}
\eea
Here $u_j^{(\ell)},~v_j^{(\ell)}$ and $w_{j}$ are arbitrary analytic functions of $x$, $y$, and $t$, while $\alpha_{\ell},~\beta_{\ell}$ and $\gamma$ are integers to be determined.\\
%\end{widetext}

\noindent \emph{Leading order analysis}: In order to identify the most dominant (leading order) terms of (\ref{pa2}), we terminate the Laurent series (\ref{ls}) for the dependent variables at the zeroth order (that is, $u^{(\ell)}=u_{0}^{(\ell)}\phi^{\alpha_{\ell}},\quad v=v_{0}^{(\ell)}\phi^{\beta_{\ell}},~ \ell=1,2,3,...,M,$ and $w=w_{0}\phi^{\gamma}$) and substitute them into Eq. (\ref{pa2}). At the leading order ($\phi^{-3}$) we get the following equations:
\bes\bea
&& w_0=2 \phi_x^2, \qquad \mbox{($2M$ times)}\\
&& \phi_x\phi_t=\sum_{\ell=1}^M c_{\ell} u_0^{(\ell)}v_0^{(\ell)},
\eea\label{mloe2}\ees
with the following condition on the leading order coefficients
\bea && \alpha_{\ell}=\beta_{\ell}=-1, \quad \ell=1,2,...,M, \qquad \gamma=-2. \label{mloe3}\eea
\emph{Resonances}: We obtain the following resonance equation (expressed below in terms of block matrices) at the order $\phi^{j-3}$, after substituting Eqns. (\ref{ls}) and (\ref{mloe3}) into Eq. (\ref{pa2}).
\bea
\left(\begin{array}{cc}
\mathbb{A} & \mathbb{P} \\
\mathbb{B}& (2-j)\phi_{t}\\
\end{array}\right) \left(\begin{array}{c}
                     \mathbb{Q} \\
                      w_j
                   \end{array}\right)
={\bf 0}, \label{resmat}
\eea
where the block matrices $\mathbb{A}$, $\mathbb{P}$, $\mathbb{B}$ and $\mathbb{Q}$ of dimensions ($2M\times 2M$), ($2M\times 1$), ($1\times 2M$) and ($2M\times 1$), respectively, are defined as $\mathbb{A}=-j(j-3)\phi_{x}^2 \mathbb{I}$, $\mathbb{B}=2(j-2)\phi_{x}(c_1v_0^{(1)}, c_1u_0^{(1)}, c_2v_0^{(2)}, c_2u_0^{(2)},...,c_{M} v_0^{(M)},~c_{M} u_0^{(M)})$, $\mathbb{P}=(u_0^{(1)}, v_0^{(1)}, u_0^{(2)}, v_0^{(2)},..., u_0^{(M)}, v_0^{(M)})^T$ and $\mathbb{Q}=(u_j^{(1)}, v_j^{(1)}, u_j^{(2)}, v_j^{(2)},..., u_j^{(M)}, v_j^{(M)})^T$. Here, $\mathbb{I}$ is a ($2M\times 2M$) identity matrix and `$T$' appearing in the superscript represents the transpose of matrix. The resonances of (\ref{pa2}) are obtained from the resonance equation (\ref{resmat}) as
\bea
j=-1,\underbrace{0,\cdots,~0}_{(2M-1)},~2,\underbrace{3,\cdots,~3}_{(2M-1)},~4.
\eea
\emph{Arbitrary analysis}: Since all the obtained resonances are integers, the system (\ref{model}) will be integrable if it possesses sufficient number of arbitrary functions. Obviously, the resonance $j=-1$ corresponds to the arbitrariness of the non-characteristic manifold $\phi(x,y,t)$. One can also explicitly prove the existence of sufficient number of arbitrary parameters at each resonance value with the help of symbolic computation for arbitrary $M$-component case by expanding the Laurent series (\ref{ls}) up to the maximum resonance value ($j=4$). At the coefficient of $\phi^{-3}$, we get only two relations (nothing but the leading order equations (\ref{mloe2})) for ($2M+1$) number of functions, which proves the arbitrariness of ($2M-1$) number of functions at resonance $j=0$. At the coefficient of $\phi^{-2}$, we have no arbitrary functions, where we obtain ($2M+1$) distinct equations for the same number of functions. We have noticed the significance of $\delta^{(\ell)}$ parameters of the present multicomponent system (\ref{model}) during the arbitrary analysis for the resonance $j=3$. The arbitrariness at $j=3$ requires ($2M-1$) number of functions to be arbitrary (out of $2M+1$ functions). This can be obtained only when all $\delta^{(\ell)}$ values are the same, i.e., $\delta^{(1)} = \delta^{(2)} =\delta^{(3)} = ... =\delta^{(M)} \equiv \delta$. Otherwise, the considered system will have an insufficient number of arbitrary functions and for such unequal $\delta^{(\ell)}$ parameters the system (\ref{model}) becomes non-integrable. Hence we can conclude that the $M$-LSRI system (\ref{model}) is Painlev\'e integrable for arbitrary nonlinearity coefficients $c_{\ell},~\ell=1,2,3,...,M$, which can admit any real values and for equal $\delta^{(\ell)}$ values.

\end{document}